\documentclass[12pt,preprint]{aastex}

\slugcomment {}

\shorttitle{Discovery of the Young L Dwarf WISE 1741$-$4642}
\shortauthors{Schneider et al.}

\begin{document}

\title{Discovery of the Young L Dwarf WISE J174102.78$-$464225.5}

\author{Adam C. Schneider\altaffilmark{a}, Michael C. Cushing\altaffilmark{a}, J. Davy Kirkpatrick\altaffilmark{b}, Gregory N.\ Mace\altaffilmark{b,c}, Christopher R. Gelino\altaffilmark{b,d}, Jacqueline K. Faherty\altaffilmark{e}, Sergio Fajardo-Acosta\altaffilmark{b}, and Scott S. Sheppard\altaffilmark{f}}

\altaffiltext{a}{Department of Physics and Astronomy, University of Toledo, 2801 W. Bancroft St., Toledo, OH 43606, USA; Adam.Schneider@Utoledo.edu}
\altaffiltext{b}{Infrared Processing and Analysis Center, MS 100-22, California Institute of Technology, Pasadena, CA 91125, USA}
\altaffiltext{c}{Department of Physics and Astronomy, UCLA, 430 Portola Plaza, Box 951547, Los Angeles, CA 90095-1547, USA}
\altaffiltext{d}{NASA Exoplanet Science Institute, Mail Code 100-22, California Institute of Technology, 770 South Wilson Ave, Pasadena, CA 91125, USA}
\altaffiltext{e}{Department of Astronomy, University of Chile, Camino El Observatorio 1515, Casilla 36-D, Santiago, Chile}
\altaffiltext{f}{Department of Terrestrial Magnetism, Carnegie Institution of Washington, 5241 Broad Branch Rd. NW, Washington, DC 20015, USA}

\begin{abstract}
\end{abstract}

We report the discovery of the L dwarf WISE J174102.78$-$464225.5, that was discovered as part of a search for nearby L dwarfs using the Wide-field Infrared Survey Explorer (WISE).  The distinct triangular peak of the H-band portion of its near-infrared spectrum and its red near-infrared colors ($J-K_S$ = 2.35$\pm$0.08 mag) are indicative of a young age.  Via comparison to spectral standards and other red L dwarfs, we estimate a near-infrared spectral type to be L7$\pm$2 (pec).  From a comparison to spectral and low-mass evolutionary models, we determine self-consistent effective temperature, log $g$, age, and mass values of 1450$\pm$100 K, 4.0$\pm$0.25 (cm s$^{-2}$), 10-100 Myr, and 4-21 M$_{Jup}$, respectively.  With an estimated distance of 10-30 pc, we explore the possibility that WISE J174102.78-464225.5 belongs to one of the young nearby moving groups via a kinematic analysis and we find potential membership in the $\beta$ Pictoris or AB Doradus associations.  A trigonometric parallax measurement and a precise radial velocity can help to secure its membership in either of these groups.  


\section{Introduction}
The {\it Wide-field Infrared Survey Explorer} (WISE) has heralded in an unprecedented era of substellar discovery.  WISE surveyed the entire sky at 3.4, 4.6, 12, and 22 $\mu$m (W1, W2, W3, and W4 bands), with channels W1 and W2 designed specifically to be sensitive to cold brown dwarfs with effective temperatures ($T_\mathrm{eff}$) less than 1400 K \citep{wri10}. With the substantial sample of substellar objects discovered by WISE (e.g. \citealt{kirk11}, \citealt{mace13}, and \citealt{thom13}), we can now perform individual in-depth studies to probe the evolution of such objects.  It is critical when investigating the evolution of substellar objects to identify sources at various ages in order to construct a complete evolutionary sequence.  The majority of L and T dwarfs discovered to date are old (age $>$ 1 Gyr), consistent with a field population (\citealt{dahn02}, \citealt{schmidt07}, \citealt{fah09}).  Young brown dwarfs are rarer, but have distinctive characteristics that distinguish them from their older counterparts.  Since young dwarfs are still contracting to their final radii, their surface gravities are lower, which can produce noticeable effects in the optical and near-infrared spectra of substellar objects (e.g. \citealt{kirk06}, \citealt{cruz09}, \citealt{bonn13}, \citealt{allers13}).  Consequently, many young substellar objects have been identified by their low surface gravity spectral features, such as enhanced VO bands, narrow or weak alkali lines, a triangular H-band spectral shape, and red near-infrared colors (\citealt{cruz07}, \citealt{kirk08}, \citealt{cruz09}, \citealt{kirk10}, \citealt{fah12}, and \citealt{thom13}). Studying these young field L and T dwarfs will help to constrain and improve substellar evolution models, while additional examples of young, late-type substellar objects can help to fill in the evolutionary gaps of these astronomical bodies.  These models will become increasingly important in interpreting results from the next generation of extreme adaptive optics imagers (i.e. the Gemini Planet Imager (GPI), Project 1640, the Spectro-Polarimetric High-contrast Exoplanet REsearch (SPHERE), the Coronagraphic High Angular Resolution Imaging Spectrograph (CHARIS)) as they attempt to image young exoplanets directly, since the determination of an accurate age can significantly affect the mass estimates of brown dwarfs and exoplanets (see, for example, the discussion of HR 8799 b, c, d, and e in \citealt{opp13}).  

Here we report the discovery of the young, late-type object WISE J174102.78$-$464225.5 (WISE 1741$-$4642, henceforth), discovered as part of a search for bright L and T type objects using WISE data.  We summarize the identification of WISE 1741$-$4642 and our subsequent observations in Section 2.  In Section 3 we discuss the spectral and physical characteristics of WISE 1741$-$4642, and provide a kinematic analysis where we evaluate potential young moving group membership for this object.              

\section{Identification and Observations}

\subsection{Identification}
WISE 1741$-$4642 was discovered as part of a search for bright L and T dwarfs.  A complete description of the selection criteria is outlined in Section 2.1 of \cite{thom13}.  To summarize, candidates are required to have a $W1 - W2$ color $>$ 0.4 and a W2 signal to noise ratio (SNR) $>$ 20.  In an attempt to exclude extragalactic objects, candidates are also required to have $W1 - W2$ $>$ 0.96($W2 - W3$) $-$ 0.96.  Candidates are required to have no matching 2MASS or USNO source within 3\arcsec to ensure moderate to high proper motions.  Lastly, candidates are required to not be flagged as a solar system object.  This search has resulted in the detection of 55 L and T dwarfs, previously discussed in \cite{kirk11}, \cite{mace13}, and \cite{thom13}. 

\subsection{Observations}

\subsubsection{IRTF/SpeX}

We observed WISE 1741$-$4642 on the night of 2012 July 19 UT employing the medium-resolution infrared spectrograph SpeX \citep{rayner2003} at the 3~m NASA Infrared Telescope Facility (IRTF) on Mauna Kea. Observations were conducted in prism mode with a 0$\farcs$5 wide slit, which achieves a resolving power ($\lambda$/$\Delta\lambda$) of $\sim$150 over the range 0.8 - 2.5 $\mu$m.  A series of 10 120s exposures were obtained at two different nod positions along the 15$\arcsec$ long slit, which was oriented to the parallactic angle. The A0V star HD~161706 was observed at a similar airmass as WISE 1741$-$4642 for telluric correction and flux calibration.  Calibration spectra of internal flat field and argon arc lamps were obtained for flat fielding and wavelength calibration. The data were reduced with Spextool, the IDL-based data reduction package for SpeX \citep{cushing2004}, which first corrected raw frames for non-linearity then pair-subtracted and flat fielded. The spectra were then optimally extracted \citep[e.g.,][]{horne1986} and wavelength calibrated using argon lamp exposures. Finally, multiple spectra were averaged together, corrected for telluric absorption and flux calibrated \citep{vacca2003}.  The spectrum was absolutely flux calibrated using 2MASS magnitudes and the procedure outlined in \cite{cush05}.

\subsubsection{Magellan/FIRE}
WISE 1741$-$4642 was observed with the Folded-port Infrared Echellette (FIRE; \citealt{simcoe13}) on the Magellan Baade Telescope on 2013 August 9 and 2013 August 10 UT in cross-dispersed echelle mode ($\lambda$/$\Delta\lambda$ $\approx$ 6000) from 0.65 - 2.55 $\mu$m.  Four 500s exposures were taken each night in an ABBA dither pattern.  The A0V stars HIP 84305 (V = 8.41) and HIP 89666 (V = 8.65) were also observed for telluric correction purposes.  We also obtained flat-field and arc lamp (ThAr) exposures for pixel response and wavelength calibration.  Data were reduced using the FIREHOSE package, which employs modified routines from the SpeXtool data reduction package (\citealt{vacca2003}, \citealt{cushing2004}).             

\section{Discussion}

\subsection{Comparison to Spectral Standards}
The SpeX near-infrared spectrum is presented in Figure 1, with prominent atomic and molecular lines labeled. We first attempted to determine the spectral type of WISE 1741$-$4642 by comparing the J-band spectral morphology with that of the near-infrared spectral standards (from \citealt{kirk10} - obtained from the Spex Prism Spectral Library\footnote{http://pono.ucsd.edu/$\sim$adam/browndwarfs/spexprism/library.html}).  A comparison of WISE 1741$-$4642 to several spectral standards is shown in Figure 2. The figure makes clear that WISE 1741$-$4642 is redder than normal L dwarfs, and a single well-fitting spectral type is not easily determined by direct comparison (see \citealt{mace13} for an in depth discussion on the difficulties encountered when comparing red dwarfs to spectral standards). We conservatively estimate the spectral type of WISE 1741$-$4642 to be L7$\pm$2 from the comparison to L and T dwarf spectral standards based on H$_2$O and FeH band strengths.

We also investigated the near-infrared spectral type indices examined by \cite{allers13}.  Most of these indices are degenerate for spectral types later than L5.  Because the by-eye comparisons suggest a spectral type later than L5, only the H$_2$OD index, originally defined by \cite{mcl03}, is useful.  We calculated an H$_2$OD index of 0.78 which,  utilizing the polynomial fits in Table 3 of \cite{allers13}, gives a spectral type of $\sim$L5.5.                  

\subsection{Comparison to Red L Dwarfs}
The spectrum of WISE 1741$-$4642 was compared to other unusually red late-type L dwarfs to aid in constraining its spectral type.  Figure 3 shows the near-infrared SpeX spectrum of WISE 1741$-$4642 compared with the red L dwarf 2MASSW J224431.67$+$204343.3 (2M2244+2043; SpT (NIR) = L7.5 (pec) - \citealt{kirk08}, SpT = L6 (VL-G) \citealt{allers13}), the young L dwarf 2MASS J035523.27$+$113343.7 (2M0355+1133; SpT (opt) = L5$\gamma$ - \citealt{fah13}), and the red L dwarf WISEP J004701.06$+$680352.1 (W0047$+$6803; SpT (NIR) = L7.5 (pec) - \citealt{gizis12}).  We derive a near-infrared spectral type of L7 (pec) for 2M0355+1133 using the method outlined in Appendix A.  All three spectra match WISE 1741$-$4642 reasonably well, suggesting a spectral type range of L7 - L8.  While these red L dwarfs are much better comparisons than the L dwarf spectral standards, differences can still be seen.  Figure 3 shows that WISE 1741$-$4642 is both redder than 2M0355$+$1133 and 2M2244$+$2043 and has a more peaked H-band profile then 2M2244$+$2043.  WISE 1741$-$4642 also shows deeper H$_2$O absorption than 2M0355$+$1133.  W0047$+$6803 is the best match at J, though is slightly redder than WISE 1741$-$4642.  A more detailed comparison of individual spectral lines is performed in Section 3.4.2.  Combining the estimated spectral type of WISE 1741$-$4642 from the comparison of spectral standards, the H$_2$OD spectral type index, and the comparison to other red L dwarfs, we assign a spectral type of L7$\pm$2 (pec) to WISE 1741$-$4642. 

\subsection{Comparison to Model Spectra}
The ``BT-Settl'' models from the Phoenix/NextGen group \citep{allard11}, using the solar abundances from \cite{asp09}, were used to estimate the $T_\mathrm{eff}$ and log $g$ values of WISE 1741$-$4642. We compared models from 1000 to 2000 K in steps of 100K and log $g$ values of 3.5 to 5.5 (cm s$^{-2}$) in steps of 0.5 dex to our measured SpeX spectrum.  In order to compare the models to our measured spectrum, each model spectrum was resampled to be uniform in ln($\lambda$) space, smoothed with a Gaussian kernel to a resolving power of 150, and resampled onto the same wavelength grid as the SpeX spectrum.  Following \cite{cush08}, a goodness of fit parameter is evaluated for each model spectrum.  The goodness of fit parameter is defined as

\begin{equation}
G_{k} = \sum\limits_{i=1}^n \left(\frac{f_i - C_k\mathcal{F}_{k,i}}{\sigma_i}\right)^2
\end{equation}

\noindent where $n$ corresponds to each data pixel, $f_i$ is the flux density of the data, $\mathcal{F}_{k,i}$ is the flux density of the model $k$, and $\sigma_i$ are the errors for each observed flux density.  The $C_k$ parameter is a multiplicative constant given by

\begin{equation}
C_{k} = \frac{\sum f_i\mathcal{F}_{k,i}/\sigma_i^2}{\sum \mathcal{F}_{k,i}/\sigma_i^2}.
\end{equation}

\noindent $C_k$ is equivalent to $(R/d)^2$, where $R$ is the stellar radius and $d$ is the distance (see Section 3.4.1).  We performed a Monte Carlo simulation using our observed spectrum to determine the uncertainty of the best fitting spectrum (e.g.\ \citealt{cush10}).  We generated 1000 simulated data sets using both the absolute flux calibration uncertainty and the individual flux uncertainties at each wavelength.  Each measured flux in the SpeX spectrum is randomly drawn from a single Gaussian distribution centered on the observed flux with a width given by the observed variance.  The entire spectrum is also multiplied by a factor randomly drawn from a Gaussian distribution centered at one with a width given by the absolute flux calibration uncertainty.  

Several low-gravity spectral models provided the best fits (Figure 4). The best fitting model spectrum was that with $T_\mathrm{eff}$ = 1600 K and a log $g$ value of 3.5.  In Section 3.4.3 we show that a $T_\mathrm{eff}$ of 1600 K and a log $g$ of 3.5 suggest an unphysical age of $\leq$ 5 Myr for WISE 1741$-$4642.  Therefore, we selected the next two best fitting spectral models ($T_\mathrm{eff}$ = 1500 K and log $g$ = 4.0; $T_\mathrm{eff}$ = 1400 K and log $g$ = 4.0) for further analysis.  The three best fitting models are displayed in Figure 4.  We take $T_\mathrm{eff}$ = 1450$\pm$100 K and log $g$ = 4.0$\pm$0.25 (cm s$^{-2}$) as our final estimates for these parameters.  This temperature is in agreement with temperatures deduced for other $\sim$L7 dwarfs \citep{gol04}.     

\subsection{Physical Characteristics}
\subsubsection{Distance}
A byproduct of the model fitting technique is an estimate of the multiplicative constant $C_k$ that is equivalent to ($R$/$d$)$^2$, where $R$ is the radius of the star and $d$ is the distance from Earth.  Therefore, a stellar radius, in combination with model spectrum comparisons, can provide a distance estimate.  We utilized four sets of low-mass evolutionary models combined with our best fit $T_\mathrm{eff}$ and log $g$ ranges to estimate a stellar radii;  the ``COND'' models \citep{bar03}, the ``DUSTY'' models \citep{cha00}, and the cloudy and cloudless models of \cite{sau08}.  For the ``DUSTY'' models, dust grains are suspended, while in the ``COND'' models dust is removed from the photosphere.  Similarly, the cloudy and cloudless models of \cite{sau08} account for cloud sedimentation, where the sedimentation efficiency parameter ($f_{sed}$) = 2 for the cloudy models and the cloudless models include no opacity from clouds. We found a radius range of 0.137-0.161 R$_{\sun}$ (1.36-1.60 R$_{Jup}$) for the ``DUSTY'' and ``COND'' models, and a range of 0.1398-0.1752 R$_{\sun}$ (1.39-1.74 R$_{Jup}$) for the \cite{sau08} models. This range, in combination with the $C_k$ values determined via model fitting result in a so-called ``spectroscopic parallax'' \citep{bowler09} of 8-11 pc.            

We also estimate photometric distances using the relations of \cite{loop08}, \cite{kirk11}, and \cite{dup12}.  For a spectral type range of L5-L9 and the 2MASS and WISE magnitudes given in Table 1, we calculate distance ranges of 12-25 pc and 11-28 pc for the \cite{loop08} and \cite{dup12} relations, respectively.   For the WISE W2 relation of \cite{kirk11}, we calculated a distance range of 11-14 pc.  The distance estimates using the $K_S$, W1, and W2 magnitudes provided the closest matches to the distance derived via spectroscopic parallax (11-20 pc, 11-17 pc, and 10-15 pc for $K_S$, W1, and W2, respectively), which could imply that the magnitudes from the wavelength region between $K_S$ and WISE W2 is the least affected by youth.  This is further supported by Figure 9 of \cite{fah13}, which shows that the absolute flux of 2M0355$+$1133 matches the field L5 2M1507-16 well at K, while it is underluminous everywhere blueward of $\sim$2 $\mu$m.  \cite{liu13} also find that many of their young mid-L dwarfs with measured parallaxes are fainter at J than field objects of comparable spectral type.  This is supported by the results of \cite{fah12}, in which they find that $\sim$80\% of objects in their low-gravity and young companion L dwarf sample are underluminous at J.  According to Table 10 of \cite{fah12}, low-gravity and young companions are, on average, 0.48 mag underluminous at J and only 0.17 mag underluminous at K.  If WISE 1741$-$4642 is similarly underluminous at J, then one would expect the distance estimates based on its J-band magnitude to be overestimated.  We found that the J-band photometric distances were the largest of all estimates; 17-28 pc and 16-26 for spectral types L5-L9 according to the \cite{dup12} and \cite{loop08} relations, respectively.  We conservatively take 10-30 pc as our final distance estimate.      


\subsubsection{Empirical Age}
The near-infrared spectrum of WISE 1741$-$4642 shows a distinctly peaked H-band continuum, a well known characteristic of young brown dwarfs (\citealt{lucas01}; Figures 1-3).  The shape of the H-band continuum is determined by differing sources of opacity production that depend on an object's surface gravity (\citealt{rice11}, \citealt{bar11}, \citealt{marley12}).  For higher gravities, the H-band shape is dominated by collision-induced absorption (CIA) opacity or cloud opacity, which results in a rounded appearance.  For a low surface gravity object, cloud opacity and CIA become less significant, and the H-band shape is governed by molecular bands around 1.7 $\mu$m, most notably H$_2$O, which results in a peaked triangular appearance.  

WISE 1741$-$4642 also shows a redder $J-K_S$ color ($J-K_S$ = 2.35$\pm$0.08 mag) and $W1-W2$ color ($W1-W2$ = 0.63$\pm$0.03 mag) compared to objects of similar spectral type (Figures 2 and 5).  \cite{fah13} demonstrate that the $J-K_S$ and $W1-W2$ colors of low surface gravity dwarfs deviate from the mean population. The triangular H-band shape and red $J-K_S$ and $W1-W2$ colors of WISE 1741$-$4642 are features typical of young dwarfs and are often caused by a low surface gravity.  \cite{kirk08} show that gravity-sensitive spectral features are only evident in objects with Pleiades ages or younger.  Therefore, the ability to distinguish low-gravity spectral features implies an age $<$ 100 Myr.  Following the example of \cite{cruz09}, we take 10 Myr as a lower age limit because WISE 1741$-$4642 is isolated from young, dense, star forming regions.  This age is also an approximate lower limit for the ages of young nearby moving groups.  

\cite{allers13} define several gravity sensitive spectral indices that take advantage of gravity sensitive of FeH bands (centered on the 0.99 and 1.20 $\mu$m bandheads), VO absorption (centered on the 1.06 $\mu$m bandhead), the K I line at 1.14 $\mu$m, and the H-band continuum shape.  We note here that the alkali and molecular band indices are most useful for spectral types between $\sim$M9 and L5, and are degenerate for our predicted spectral type of L7.  We can, however, determine the $H$-cont index, which is designed to quantify the H-band continuum shape.  We calculated an $H$-cont value of 0.955 for our SpeX spectrum, which follows the low gravity sequence, well separated from the location of normal field dwarfs (see Figure 24 of \citealt{allers13}).  For the similarly red L7.5 (pec) dwarf W0047$+$6803, we calculate an $H$-cont value of 0.963.  We can also utilize the K I pseudo-equivalent width (EW) definitions of \cite{allers13} using our moderate-resolution FIRE spectrum.  Alkali lines are known to be gravity sensitive, and can appear narrower than those seen in normal field dwarfs because of the lack of pressure broadened wings \citep{kirk06}.  We first smoothed our FIRE spectrum with a gaussian kernel to a resolution (R $\sim$ 1500) similar to those inspected in \cite{allers13}.  We measured pseudo-EW values for the 1.169, 1.177, and 1.252 $\mu$m K I lines of 0.93, 1.89, and 0.86 \AA, respectively. We similarly measure the pseudo-EWs of W0047$+$6803 (3.31, 2.62, and 1.16 \AA) and W0355$+$1133 (2.58, 2.59, and 0.98 \AA).  All three measured pseudo-EWs are consistent with a low-gravity object (see Figure 23 and Table 10 of \citealt{allers13}).  Figures 6 and 7 display the K I spectral line regions of WISE 1741$-$4642 compared with the red L dwarfs W0047$+$6803 and W0355$+$1133.  The FIRE spectrum has been smoothed in both figures to a resolution comparable to each comparison spectrum.  Figures 6 and 7 show WISE 1741$-$4642 has comparable K I lines strengths to both W0047$+$6803 and W0355$+$1133.  We also compared the K I lines with field brown dwarfs from the NIRSPEC Brown Dwarf Spectroscopic Survey (BDSS) \citep{mcl07}.  A comparison with the L5 field dwarf 2M1507$-$16 and the L7 field dwarf DENIS 0205$-$11 are shown in Figures 8 and 9.  Each figure shows that the K I lines of WISE 1741$-$4642 are weaker and/or narrower than those lines seen in field dwarfs of a similar spectral type.  We note that K I line strengths are also sensitive to temperature differences, which may contribute to the dissimilarity of the spectra.           

\cite{canty13} show that the K-band continuum shape can easily distinguish young ($\leq$ 10 Myr) brown dwarfs from an older field population, and can be used with confidence even with low resolution spectra.  We note that the sample used in \cite{canty13} has a dearth of objects with intermediate ages ($>$10 Myr and less than the age of field dwarfs) and the effectiveness of the $H_2(K)$ index in this age range is not well constrained.  They define an $H_2(K)$ index to quantify the K-band continuum slope, given as

\begin{equation}
H_2(K) = \frac{F_\lambda(2.17\mu m)}{F_\lambda(2.24\mu m)}
\end{equation}

\noindent where $F_\lambda(2.17\mu m)$ and $F_\lambda(2.24\mu m)$ are the median fluxes over a total range of 0.02 $\mu$m centered at 2.17 and 2.24 $\mu$m, respectively.  We calculated the $H_2(K)$ index of WISE 1741$-$4642 and found a value of 1.029.  However, we note that the analysis of K-band continuum shape by \cite{canty13} for young brown dwarfs was performed for spectral types $\leq$ L0.  To extend the work of \cite{canty13} into the L dwarf regime, we gathered all available spectra from the Spex Prism Spectral Library for which a resolving power $\geq$120 is available.  Since many of the given or published spectral types were not determined homogeneously, we independently derived near-infrared spectral types for each available spectrum based on the \cite{kirk10} L dwarf classification scheme (See Appendix A). The $H_2(K)$ index of each L dwarf was calculated, and is shown in Figure 10 and given in Table 2.  Any spectrum found to be especially noisy was not included in this analysis.  While a spectral type dependance is seen past L0 (the $H_2(K)$ index generally increases with increasing spectral type), WISE 1741-44642 and the red L dwarfs 2M0355$+$1133 and 2M0047$+$6803 are seen to deviate from the field population.  Also showing strong deviations below their near-infrared spectral types are the young, red L6 dwarf 2MASS J214816.28$+$400359.3 \citep{loop08b}, the young, red L2 dwarf 2MASS J014158.23-463357.4 \citep{kirk06}, the low gravity L dwarf 2MASS J224953.45+004404.6 (SpT$_{opt}$ = L4$\gamma$; \citealt{fah13}), and the red L8 dwarf SDSSp J010752.33$+$004156.1 \citep{geb02}. 

In summary, WISE 1741$-$4642 shows several traits, such as a peaked H-band continuum shape, red near-infrared colors, and a low $H_2(K)$ index, that are characteristic of a young age.                                 
   
\subsubsection{Model Age and Mass}
We compared the position of WISE 1741$-$4642 in $T_\mathrm{eff}$ vs.\ log $g$ space with the ``COND'' \citep{bar03},  ``DUSTY'' \citep{cha00}, and \cite{sau08} cloudy and cloudless low-mass evolutionary models (discussed in Section 3.4.1).  We note here that the spectral models of \cite{allard11} and the low-mass evolutionary models used in Section 3.4 are not self-consistent.  However, such a comparison is the only available procedure for determining the physical characteristics of WISE 1741$-$4642.  Figures 11 and 12 show that the best fitting model spectrum ($T_\mathrm{eff}$ = 1600, log $g$ = 3.5) suggests an age $\leq$ 5 Myr, but there are no regions of ongoing star formation within 30 pc.  Because this is unphysical for WISE 1741$-$4642, we use the next two best fitting models.  For $T_\mathrm{eff}$ = 1450$\pm$100 K and log $g$ = 4.0$\pm$0.25, the ``COND'' and ``DUSTY'' models suggest an age between 5 and 100 Myr (Figure 11).  The \cite{sau08} models suggest an age of 3-100 Myr (Figure 12). Both model ages are consistent with the age implied by the near-infrared spectral characteristics of WISE 1741$-$4642, and we take 10-100 Myr as our final age estimate for WISE 1741$-$4642.  We take 10 Myr as a lower age limit because WISE 1741$-$4642 is isolated from young, dense, star forming regions.  Even though the best fitting model was deemed unphysical, this type of object has the potential to anchor low mass evolutionary models for young ages.    

The same models and spectral model parameters are used to estimate the mass of WISE 1741$-$4642.  The ``COND'' and ``DUSTY'' models suggest a mass range of 5-21 M$_{Jup}$, while the \cite{sau08} models suggest a mass range of 4-15 M$_{Jup}$ (Figures 11 and 12).  We take 4-21 M$_{Jup}$ as our final mass estimate.

\subsection{Kinematic Analysis of WISE 1741$-$4642}
Because WISE 1741$-$4642 is located in the southern hemisphere, where many young moving groups and associations reside, we considered the possibility that it belongs to one of these.  Although a full moving group membership analysis requires accurate parallax, proper motion,  and radial velocity measurements, we may still explore how probable nearby moving group membership is for WISE 1741$-$4642.  We calculate proper motions in right ascension and declination using the WISE All-Sky and 2MASS positions of WISE 1741$-$4642 (listed in Table 1).  Using a distance estimate of 10-30 pc and the proper motions given in Table 1, we calculated UVW\footnote{UVW space motions are calculated using the measured radial velocity. $UVW$ are defined with respect to the Sun. U is positive toward the Galactic center, V is positive in the direction of Galactic rotation, and W is positive toward the north Galactic pole.} space motions for WISE 1741$-$4642 for a range of radial velocities.  Then we compared the resulting UVW space motions to the ``good box'' from \cite{zuck04}, a region of UVW space where nearby moving group members reside.  Because UVW space motions depend on an accurate distance measurement, we calculated them using 10, 20, and 30 pc as the distance to WISE 1741$-$4642.  There are a range of acceptable radial velocities that would put WISE 1741$-$4642 in the UVW space which coincides with nearby young moving groups, illustrated in Figure 13.  Figure 13 also shows that WISE 1741$-$4642 is much more likely to belong to a nearby young moving group if its distance is 10 or 20 pc, rather than 30 pc.  As shown in Section 3.5.1, the largest photometric distance estimates come from the relations of \cite{dup12} and \cite{loop08} and the J-band magnitude of WISE 1741$-$4642, which may be underluminous. 

In addition, we also took advantage of the convergent point analysis and the Bayesian young nearby association membership probability tools developed by \cite{rod13} and \cite{malo13}, respectively.  As both authors state clearly, high probabilities do not guarantee young association membership.  However, these tools can aid in illustrating the likelihood of nearby moving group membership for WISE 1741$-$4642.  

The convergent point analysis tool of \cite{rod13} uses the position and proper motion of an astronomical body to calculate parallel and perpendicular proper motion components relative to the convergent point of several nearby young associations.  WISE 1741$-$4642 membership probabilities for the TW Hydrae Association (TWA), the Tucana-Horologium association (Tuc-Hor), the Carina-Near moving group, and the Columba moving group are all less than 10\%.  Membership probabilities for the $\beta$ Pictoris and AB Doradus associations are 94.2 and 91.9\%, respectively.  Furthermore, the convergent point analysis tool returns the ideal kinematic distances and radial velocities for membership in each group.  The kinematic distance is predicted to be 12.3 pc for $\beta$ Pictoris and 19.1 pc for AB Doradus, distances that match well with our photometric distance estimates.  The predicted radial velocities are -5.4 km s$^{-1}$ and 2.1 km s$^{-1}$ for $\beta$ Pictoris and AB Doradus, respectively.

The Bayesian Analysis for Nearby Young AssociatioNs tool (or BANYAN; \citealt{malo13}) uses Bayesian inference to determine association membership probabilities, predicted radial velocities, and the most probable distances for several young nearby associations.  Unlike the convergent point analysis tool of \cite{rod13}, the BANYAN tool determines membership probabilities so that they sum to 100\%.  Using the WISE All-sky catalog position and the proper motions provided in Table 1, the BANYAN tool determined that WISE 1741$-$4642 has a 0\% probability of belonging to TWA, Tuc-Hor, Columba, and Carina.  Membership probabilities for $\beta$ Pictoris, AB Doradus, Argus, and old field stars are 40.75\%, 43.45\%, 15.34\%, and 0.47\%, respectively (Argus is not considered by the convergent point analysis tool of \citealt{rod13}).  As with the convergent point analysis tool, the BANYAN tool claims a high likelihood that WISE 1741$-$4642 belongs to either $\beta$ Pictoris or AB Doradus ($\sim$84\% total).  Statistical distances are in excellent agreement with those from the convergent point method as well; 13.0 pc for $\beta$ Pictoris and 19.5 pc for AB Doradus, again consistent with our photometric distance estimates.  The predicted radial velocities from BANYAN are -4.7 km s$^{-1}$ and +2.57 km s$^{-1}$ for $\beta$ Pictoris and AB Doradus, respectively.

We estimated the radial velocity of WISE 1741$-$4642 from our FIRE spectrum using measured K I central wavelengths compared to the rest wavelengths found in \cite{rayner09}.  We measured the central wavelength for each of the four K I lines at 1.16934, 1.17761, 1.24356, and 1.252566 $\mu$m and from these calculated a radial velocity, corrected for the heliocentric radial velocity, of -5.7 $\pm$ 5.1 km s$^{-1}$.  While uncertainties using this method are large, a radial velocitiy significantly discrepant from those predicted for WISE 1741$-$4642 if it is a $\beta$ Pictoris or AB Doradus member could have ruled out potential membership.  Our measured radial velocity range does not rule out $\beta$ Pictoris or AB Doradus membership.  We also examined the XYZ position of WISE 1741$-$4642 using the WISE All-Sky position and our distance estimate of 10-30 pc.  We find XYZ values of (9.5, -2.7, -1.5), (19.0, -5.3, -2.9), and (28.6, -8.0, -4.4) pc for distances of 10, 20, and 30 pc, respectively.  Comparing these values to the mean XYZ positions and dispersions for $\beta$ Pictoris (9.27$\pm$31.71, -5.96$\pm$15.19, -13.59$\pm$8.22) pc and the AB Doradus association (-2.37$\pm$20.03, 1.48$\pm$18.83, -15.62$\pm$16.59) pc from \cite{malo13}, we find that the XYZ positions of WISE 1741$-$4642 are consistent with known members of the $\beta$ Pictoris and AB Doradus associations.         

Given the youthful characteristics of the near-infrared spectrum of WISE 1741$-$4642 and the good agreement of its space motion with the $\beta$ Pictoris and AB Doradus associations, we believe there is a strong possibility of moving group membership.  The position and proper motion of WISE 1741$-$4642 along with known members of $\beta$ Pictoris and AB Dordados members from \cite{sch13} is displayed in Figure 14.  Accurate parallax and radial velocity measurements will aid in clearing up any young association membership ambiguity.                                        

\section{Conclusions}
We have shown through an analysis of the near-infrared spectrum and photometric colors of WISE 1741$-$4642 that it shows characteristics indicative of an age of 10-100 Myr.  Since WISE 1741$-$4642 is young, it is less massive than the majority of objects with similar spectral types, which typically have ages greater than 1 Gyr (assuming spectral type correlates with effective temperature). We estimated a spectral type of L7$\pm$2 by a comparison of the near-infrared spectrum to spectral standards and other red L dwarfs.  We determined an effective temperature of 1450$\pm$100 K and a log $g$ of 4.0$\pm$0.25 from a comparison to spectral models.  From the $T_\mathrm{eff}$ and log $g$ estimates, we estimate the mass of WISE 1741$-$4642 to be 4-21 M$_{Jup}$.  We also explored the possibility that WISE 1741$-$4642 is a member of one of the established nearby, young moving groups or associations.  We found that membership of WISE 1741$-$4642 in the $\beta$ Pictoris or AB Doradus moving groups is a possibility deserving follow-up observations.  Future parallax and radial velocity measurements can confirm any association membership.  
 
\acknowledgments

This research has made use of the SIMBAD database and VizieR catalog access tool, operated at CDS, Strasbourg, France.  This publication makes use of data products from the Two Micron All Sky Survey, which is a joint project of the University of Massachusetts and the Infrared Processing and Analysis Center/California Institute of Technology, funded by the National Aeronautics and Space Administration and the National Science Foundation, and the {\it Wide-field Infrared Survey Explorer}, which is a joint project of the University of California, Los Angeles, and the Jet Propulsion Laboratory/California Institute of Technology, funded by the National Aeronautics and Space Administration.  This research has made use of the NASA/ IPAC Infrared Science Archive, which is operated by the Jet Propulsion Laboratory, California Institute of Technology, under contract with the National Aeronautics and Space Administration.  This research has benefitted from the SpeX Prism Spectral Libraries, maintained by Adam Burgasser at http://pono.ucsd.edu/$\sim$adam/browndwarfs/spexprism.  This research has benefitted from the M, L, T, and Y dwarf compendium housed at dwarfarchives.org.  Visiting Astronomer at the Infrared Telescope Facility, which is operated by the University of Hawaii under Cooperative Agreement no. NNX$-$08AE38A with the National Aeronautics and Space Administration, Science Mission Directorate, Planetary Astronomy Program.  The Brown Dwarf Spectroscopic Survey is hosted by UCLA and provided an essential comparison library for our moderate resolution spectroscopy.

\appendix
\section{SpeX Near-Infrared Spectral Fitting}
The SpeX Prism Spectral Library is an invaluable resource to the brown dwarf community.  We chose to utilize this resource to extend the work of \cite{canty13} regarding the H$_2$K youth index to the entire L dwarf spectral sequence.  Since many of the given or published spectral types in the Spex Prism Spectral Library were not determined homogeneously, we independently derived near-infrared spectral types for each available spectrum based on the \cite{kirk10} L dwarf classification scheme.  For each spectrum, we first normalized to 1.28 $\mu$m, then performed a $\chi^2$ fitting to the region between 0.9 and 1.4 $\mu$m.  Each spectrum was then inspected by-eye over the entire near-infrared range from 0.8 to 2.5 $\mu$m to see which subtype was most similar to the spectrum in question, with the results of the $\chi^2$ fitting used as a guide. The newly derived near-infrared spectral types are given in Table 2, along with the reference where the original near-infrared SpeX spectrum was published.  We also include previously derived optical and near infrared spectral types, when available.  Those spectra with blue or red near-infrared colors compared to their best fitting spectral type are labeled as blue or red in the notes section of Table 2.  The newly derived infrared spectral types of the sample agree to within $\pm$2 subtypes at a rate of $\sim$89\% (101/114) with previously published infrared spectral types and $\sim$93\% (103/111) with optical spectral types.  A comparison of the newly derived spectral types with previously published spectral types is shown in Figure 15.

\begin{deluxetable}{lccc}
\tablecaption{WISE 1741$-$4642 Properties}
\tablewidth{0pt}
\tablehead{
\colhead{Parameter} & \colhead{Value} & \colhead{Ref.}}
\startdata
\cutinhead{Observed Properties}
$\alpha$ (J2000) & 17:41:02.79 & 1\\ 
$\delta$ (J2000) & $-$46:42:25.55 & 1\\ 
$\mu$$_{\alpha}$cos$\delta$ (mas yr$^{-1}$) & $-$20.4$\pm$9.2 & 3\\
$\mu$$_{\delta}$ (mas yr$^{-1}$) & $-$343.0$\pm$13.7 & 3\\
$v_{rad}$ & -5.7 $\pm$ 5.1 km s$^{-1}$ & 3\\
J (mag) & 15.786$\pm$0.075 & 1\\
H (mag) & 14.534$\pm$0.054 & 1\\
K$_S$ (mag) & 13.438$\pm$0.035 & 1\\
W1(mag) & 12.301$\pm$0.025  & 2\\
W2 (mag) & 11.675$\pm$0.023 & 2\\
W3 (mag) & 11.432$\pm$0.190 & 2\\
W4 (mag) & $<$12.301 & 2\\
$J-H$ (mag) & 1.25$\pm$0.09 & 1\\
$J-K_S$ (mag) & 2.35$\pm$0.08 & 1\\
$J-W2$ (mag) & 4.11$\pm$0.08 & 1,2\\
$W1-W2$ (mag) & 0.63$\pm$0.03 & 2\\
\cutinhead{Inferred Properties}
Spectroscopic Distance & 8-11 pc & 3\\
Photometric Distance & 10-30 pc & 3\\
Spectral Type (NIR) & L7$\pm$2 (pec) & 3\\
$T_\mathrm{eff}$ & 1450$\pm$100 K  & 3\\
log $g$ & 4.0$\pm$0.25 (cm s$^{-2}$) & 3\\
Mass & 4-21 M$_{Jup}$ & 3\\
Age & 10-100 Myr & 3\\
\enddata
\\
References: (1) 2MASS catalog \citep{cut03}; (2) WISE All-Sky Catalog \citep{cut12}; (3) This paper. 
\end{deluxetable}

\begin{deluxetable}{lcccccl}
\tablecaption{SpeX Prism Spectral Library Near-Infrared L Dwarf Spectral Types}
\tablewidth{0pt}
\tablehead{
\colhead{2MASS Name} & \colhead{SpT$_{opt}$(Pub.)} & \colhead{SpT$_{NIR}$(Pub.)} & \colhead{SpT$_{IR}$(Adopted)} & \colhead{$H_2(K)$} & \colhead{Ref.} & \colhead{note}}
\startdata
2M0004$-$4044 & L5 & L4.5 & L5 & 1.051 &  1\\
2M0015$+$2959 & L7 & L7.5 (pec) & L8.5 & - & 2\\
2M0016$-$4056 & L3.5 & - & L5.5 & 1.034 & 3\\
2M0030$-$1450 & L7 & - & L6.5 & 1.052 & 3\\
2M0032$+$1410 & - & L8 & L9 & 1.199 & 3\\
2M0036$+$1821 & L3.5 & L4 & L1.5 & 1.062 & 4\\
2M0036$+$2413 & - & L5.5 & L6 & - & 5\\
2M0051$-$1544 & L3.5 & - & L6 & 1.047 & 3\\
2M0053$-$3631 & L3.5 & - & L4 & 1.072 & 3\\
2M0103$+$1935 & L6 & L7 & L7 & 1.063 & 6\\
2M0107$+$0041 & L8 & L5.5 & L8 & - & 3 & red\\
2M0130$-$4445 & - & L6 & L7.5 & - & 7\\
2M0134$+$0508 & - & L1 & L1 & 1.022 & 2\\
2M0141$-$1601 & - & L7: & L5.5 & - & 2\\
2M0141$-$4633 & L0$\gamma$ & L0 (pec) & L2 & - & 22 & red\\
2M0144$-$0716 & L5 & - & L6.5 & 1.046 & 4\\
2M0147$+$4731 & - & L1.5 & L1 & 1.024 & 2\\
2M0205$+$1251 & L5 & - & L6.5 & 1.046 & 8\\
2M0205$-$0795 & - & L2 & L2 & 1.004 & 2\\
2M0205$-$1159 & L7 & L5.5 & L9 & 1.121 & 3 \\
2M0206$+$2235 & - & L5.5 & L4.5 & - & 5 & blue\\
2M0207$+$1355 & L3 & L3 & L1.5 & - & ?\\
2M0208$+$2542 & L1 & - & L1.5 & 1.054 & 4\\
2M0208$+$2737 & L5 & - & L6 & 1.063 & 3\\
2M0227$-$1624 & L1 & - & L1.5 & 1.023 & 4 & sl. red\\
2M0228$+$2537 & L0: & L0 & L0 & - & 4\\
2M0230$-$3027 & - & L1 & L1 & 1.026 & 2\\
2M0235$-$2331 & L1 & L1 & L1.5 & 1.038 & 4\\
2M0241$-$1241 & L2: & - & L2.5 & 1.035 & 4\\
2M0255$-$4700 & L8 & L9 & L9 & 1.144 & 9\\
2M0257$-$3105 & L8 & - & L8.5 & 1.094 & 10\\
2M0300$-$2130 & - & L6 (pec) & L4.5 & - & 2\\
2M0310$-$1648 & L8 & L9 & L9.5 & 1.140 & 1 & sl. red\\
2M0318$-$3421 & L7 & - & L7 & 1.063 & 3\\
2M0320$-$0446 & M8 & L0.5 & L0 & - & 4 & sl. blue\\
2M0328$+$2302 & L8 & L9.5 & L9.5 & 1.155 & 4\\
2M0354$+$2540 & L0 & L1 & L0 & - & 23\\
2M0406$-$3812 & L0$\gamma$ & L1 (pec) & L2 & - & 2 & red?\\
2M0439$-$2353 & L5: & - & L4.5 & 1.067 & 1\\
2M0516$-$3332 & L3: & - & L6 & - & 3\\
2M0539$-$0059 & L5 & L5 & L5 & 1.089 & ? & blue\\
2M0624$-$4521 & L5: & - & L6.5 & 1.055 & 10\\
2M0652$+$4710 & L4.5 & - & L5.5 & 1.038 & 3\\
2M0654$+$6528 & - & L6 & L5.5 & - & 3\\
2M0717$+$5705 & L3 & L6.5 & L3 & 1.036 & 3 & red\\
2M0741$+$0531 & - & L1.5 & L1 & 1.023 & 2\\
2M0801$+$4628 & L6.5 & - & L5 & 1.038 & 3\\
2M0805$+$4812 & L4 & L9 & L5 & 1.098 & 11 & blue\\
2M0820$+$1037 & - & L9.5 & L9.5 & - & 5\\
2M0820$+$4500 & L5 & - & L7 & 1.052 & 3\\
2M0823$+$2428 & L3 & - & L2 & 1.033 & 3 & sl. red\\
2M0825$+$2115 & L7.5 & L6 & L7.5 & 1.079 & 3\\
2M0828$-$1309 & L2 & - & L1 & 1.029 & 12\\
2M0830$+$4828 & L8 & L9 & L9.5 & 1.120 & 4\\
2M0835$+$1953 & - & L5 & L5 & 1.080 & 5\\
2M0835$-$0819 & L5 & - & L6.5 & 1.048 & 3\\
2M0847$-$1532 & L2 & - & L1.5 & 1.049 & 13\\
2M0850$+$1057 & L6 & - & L6.5 & 1.059 & 14\\
2M0851$+$1817 & - & L4.5 & L6 & - & 5\\
2M0852$+$4720 & - & L9.5 & L9.5 & 1.128 & 3\\
2M0856$+$2235 & L3: & - & L3 & 1.033 & 3\\
2M0857$+$5708 & L8 & L8 & L7 & 1.097 & 3 & red\\
2M0859$-$1949 & L6:: & - & L8 & 1.146 & 10\\
2M0905$+$5623 & L5 & - & L5 & 1.070 & 3\\
2M0908$+$5032 & L5 & L9 & L8 & 1.124 & 10 & blue\\
2M0915$+$0422 & L7 & - & L6 & 1.068 & 1\\
2M0921$-$2104 & L1.5 & L4 & L1 & 1.043 & 12 & blue\\
2M0929$+$3429 & L8 & L7.5 & L7 & 1.080 & 3\\
2M1007$+$1930 & - & L8 & L9 & 1.144 & 3\\
2M1010$-$0406 & L6 & L6 & L6 & 1.071 & 8\\
2M1025$+$3212 & - & L7.5 & L9 & - & 5\\
2M1022$+$4114 & L0 & L (early)& L0.5 & - & 4\\
2M1036$-$3441 & L6 & - & L9 & 1.126 & 3\\
2M1043$+$1213 & - & L7 & L9 & 1.156 & 3\\
2M1043$+$2225 & L8 & - & L8.5 & 1.152 & 10 & sl. red\\
2M1044$+$0429 & - & L7 & L8.5 & 1.130 & 3\\
2M1046$+$4441 & - & L5 (pec) & L4 & 1.096 & 2\\
2M1048$+$0111 & L1 & L4 & L1 & 1.021 & 4\\
2M1058$-$1548 & L3 & L3 & L2.5 & 1.037 & 3\\
2M1100$+$4957 & L3.5 & - & L3 & 1.052 & 10 & sl. red\\
2M1104$+$1959 & L4 & - & L4 & 1.086 & 15\\
2M1112$+$3548 & L4.5 & - & L5 & 1.042 & 3\\
2M1113$+$3430 & - & L3 & L3 & - & 5\\
2M1121$+$4332 & - & L7.5 & L5 & - & 5 & blue\\
2M1126$-$5003 & L4.5 & L6.5 (pec) & L5 & 1.102 & 16 & blue\\
2M1146$+$2230 & L3 & - & L2.5 & 1.041 & 3\\
2M1155$+$0559 & - & L7.5 & L5 & 1.078 & 3\\
2M1207$-$3151 & L3: & - & L2 & 1.036 & 3 & red\\
2M1216$+$3003 & - & L3.5 & L4 & - & 5\\
2M1219$-$3128 & - & L8 & L9 & 1.161 & 3 & sl. red\\
2M1221$+$0257 & L0 & - & L0.5 & - & 4\\
2M1228$-$1547 & L5 & L6 & L4.5 & 1.073 & 3\\
2M1230$+$2827 & - & L4: & L2 & 1.043 & 17\\
2M1239$+$5515 & L5 & - & L5 & 1.048 & 3\\
2M1247$-$1117 & - & L0: & L0 & - & 2\\
2M1302$-$5650 & L2 & L3 (pec) & L3 & - & 2\\
2M1305$-$2541 & L2 & L2 & L2 & 1.021 & 12\\
2M1318$+$1736 & - & L5.5 & L6 & 1.039 & 2\\
2M1326$-$0038 & L8: & L5.5 & L7 & 1.046 & 3\\
2M1331$+$3407 & L0 & L1 (pec) & L1 & 1.019 & 2 & sl. red\\
2M1331$-$0116 & L6 & L8 & T0 & - & 3\\
2M1342$+$1340 & - & L5.5 & L6 & - & 5\\
2M1345$+$5216 & - & L3.5 & L3 & - & 5\\
2M1357$+$1428 & L4: & - & L2 & 1.014 & 3\\
2M1400$+$4338 & - & L7 & L8 & 1.138 & 3\\
2M1407$+$1241 & L1:: & - & L4 & 1.042 & 3\\
2M1416$+$1348 & L6 & - & L5 & 1.111 & 18 & blue\\
2M1416$+$5006 & - & L5.5 & L4.5 & - & 5\\
2M1418$-$3538 & - & L1.5 & L1.5 & 1.030 & 2\\
2M1419$+$5919 & - & L1: & L1 & - & 2\\
2M1422$+$2215 & - & L6.5 & L5 & - & 5 & blue\\
2M1431$+$1436 & L2 & L3.5 & L1 & 1.018 & 17 & blue\\
2M1434$+$2202 & - & L2.5 & L1 & 1.045 & 17 & blue\\
2M1439$+$1929 & L1 & - & L1 & 1.016 & 15\\
2M1440$-$1303 & L1 & L1 (pec) & L1 & 1.038 & 2\\
2M1448$+$1031 & L4: & L3.5 & L5.5 & 1.065 & 3\\
2M1506$+$1321 & L3 & L3 & L3 & 1.020 & 9\\
2M1515$+$4436 & - & L7.5 & L7.5 & 1.093 & 3\\
2M1523$+$3014 & L8 & L8 & L8 & 1.126 & 3\\
2M1526$+$2043 & L7 & - & L5 & 1.076 & 15\\
2M1534$+$1219 & - & L4 & L3 & 1.067 & 5\\
2M1540$+$3742 & - & L9 & L9 & 1.135 & 3\\
2M1548$+$1722 & - & L5 & L4.5 & - & 5\\
2M1550$+$1455 & L2: & - & L2.5 & 1.039 & 19\\
2M1617$+$4019 & - & L4 & L6.5 & 1.064 & 5 & sl. red\\
2M1622$+$1159 & - & L6 & L4.5 & - & 5\\
2M1630$+$4344 & - & L7 & L8.5 & 1.105 & 3\\
2M1632$+$1904 & L8 & L8 & L8 & 1.078 & 15\\
2M1633$-$0640 & - & L6 & L5.5 & 1.085 & 5 & sl. blue\\
2M1638$+$1327 & - & L2 & L1.5 & 1.039 & 2\\
2M1645$-$1319 & L1.5 & - & L1 & 1.018 & 4\\
2M1649$+$4643 & - & L5 & L4.5 & - & 5 & sl. blue\\
2M1653$+$6231 & L3 & - & L1.5 & 1.028 & 3\\
2M1705$-$0516 & - & L4 & L1 & 1.034 & 3\\
2M1707$-$0558 & - & L3 & L2 & 1.067 & 13\\
2M1711$+$2232 & L6.5 & - & L9.5 & 1.137 & 3\\
2M1728$+$3948 & L7 & - & L6.5 & 1.080 & 14\\
2M1731$+$5310 & - & L6 & L5 & 1.078 & 5\\
2M1732$+$2656 & - & L1 & L1 & 1.026 & 2\\
2M1739$+$2454 & - & L4 & L2 & 1.048 & 2\\
2M1750$-$0016 & - & L5.5 & L4.5 & 1.060 & 3 & sl. blue\\
2M1807$+$5015 & L1.5 & L1 & L1 & 1.029 & 4\\
2M1813$+$5101 & - & L5 & L4.5 & 1.094 & 2\\
2M1821$+$1414 & L4.5 & L5 (pec) & L5 & 1.047 & 20\\
2M1916$-$3700 & - & L1 & L1 & - & 2\\
2M1928$-$4356 & L4 & - & L6 & 1.057 & 3\\
2M1941$+$6826 & - & L2 & L2 & 1.046 & 2\\
2M1949$+$6222 & L2 (pec) & L2 (pec) & L2 & - & 2\\
2M2002$-$0521 & L6 & - & L5.5 & 1.050 & 4 & sl. red\\
2M2026$-$2943 & L1: & - & L1 & - & 21\\
2M2028$+$0052 & L3 & - & L2.5 & 1.048 & 15\\
2M2034$+$0827 & L1 & - & L1 & 1.038 & 3\\
2M2036$+$1051 & L3 & - & L2 & 1.041 & 3\\
2M2043$-$1551 & - & L9 & L9.5 & 1.201 & 3\\
2M2057$-$0252 & L1.5 & L1.5 & L1 & 1.028 & 15\\
2M2101$+$1756 & L7.5 & - & L7.5 & 1.097 & 3\\
2M2104$-$1037 & L2.5 & - & L2 & 1.038 & 3\\
2M2130$-$0845 & L1.5 & - & L1 & 1.035 & 2\\
2M2131$-$0119 & - & L9 & L9.5 & - & 5\\
2M2132$+$1029 & - & L4.5 & L5.5 & - & 5\\
2M2132$+$1341 & L6 & - & L6 & 1.068 & 10\\
2M2133$+$1018 & - & L5 & L3.5 & - & 5\\
2M2142$-$3101 & L3 & - & L2 & 1.048 & 3\\
2M2148$+$4003 & L6 & L6 (pec) & L7 & - & 20 & red\\
2M2151$+$3402 & - & L7 (pec) & L6 & 1.084 & 2\\
2M2151$-$2441 & L3 & - & L3 & 1.034 & 4 & red\\
2M2155$+$2345 & - & L2 & L2 & 1.080 & 2\\
2M2158$-$1550 & L4: & L4 & L4 & 1.065 & 2\\
2M2211$+$6856 & - & L2 & L2 & - & 2\\
2M2212$+$3430 & L5: & L6 & L6 & 1.073 & 3\\
2M2249$+$0044 & L3 & L5 & L7 & 1.038 & 4\\
2M2252$-$1730 & - & L7.5 & T0 & - & 8\\
2M2317$-$4838 & L4 (pec) & L6.5 (pec) & L5 & 1.046 & 2 & red\\
2M2328$-$1038 & - & L3.5 & L2 & - & 5\\
2M2351$+$3010 & L5.5 & L5 (pec) & L6.5 & 1.078 & 2\\
\enddata
\\
References: (1) \citealt{burg07}; (2) \citealt{kirk10}; (3) \citealt{burg10}; (4) \citealt{burg08b}; (5) \citealt{chiu06}; (6) \citealt{cruz04}; (7) \citealt{dhit11}; (8) \citealt{reid06}; (9) \citealt{burg07}; (10) \citealt{sieg07}; (11) \citealt{burg07c}; (12) \citealt{burg07a}; (13) \citealt{mcel06}; (14) \citealt{burg11}; (15) \citealt{burg04}; (16) \citealt{burg08a}; (17) \citealt{shep09}; (18) \citealt{schmidt10}; (19) \citealt{burg09}; (20) \citealt{loop08b}; (21) \citealt{gel10}; (22) \citealt{kirk06}; (23) \citealt{burg06b}                
\end{deluxetable}

\begin{figure}
\plotone{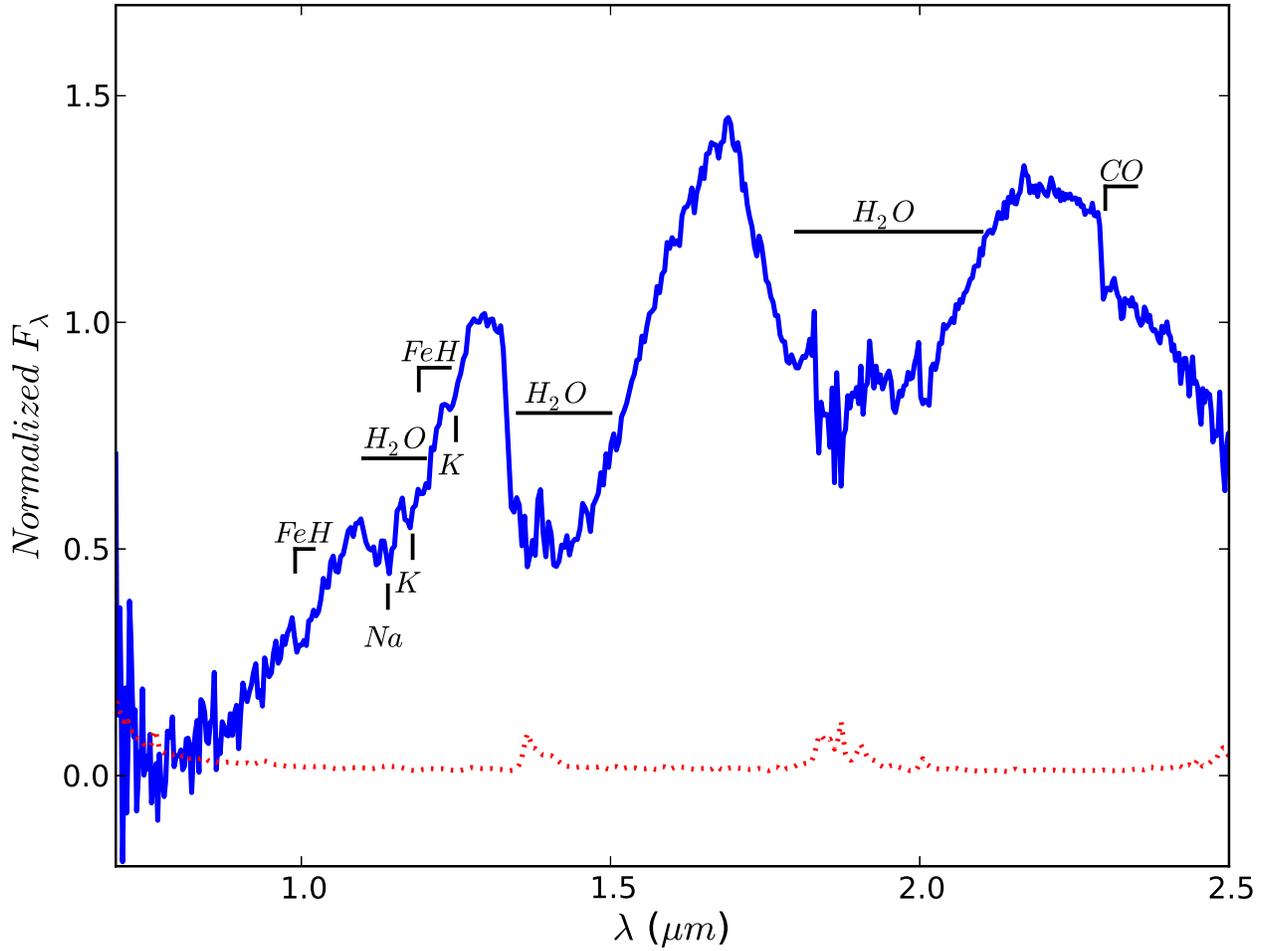}
\caption{The SpeX near-infrared spectrum of WISE 1741$-$4642.  Uncertainties are shown by a red dotted line.  Major atomic and molecular features identifiable at low resolution have been labeled.}
\end{figure}

\begin{figure}
\epsscale{0.9}
\plotone{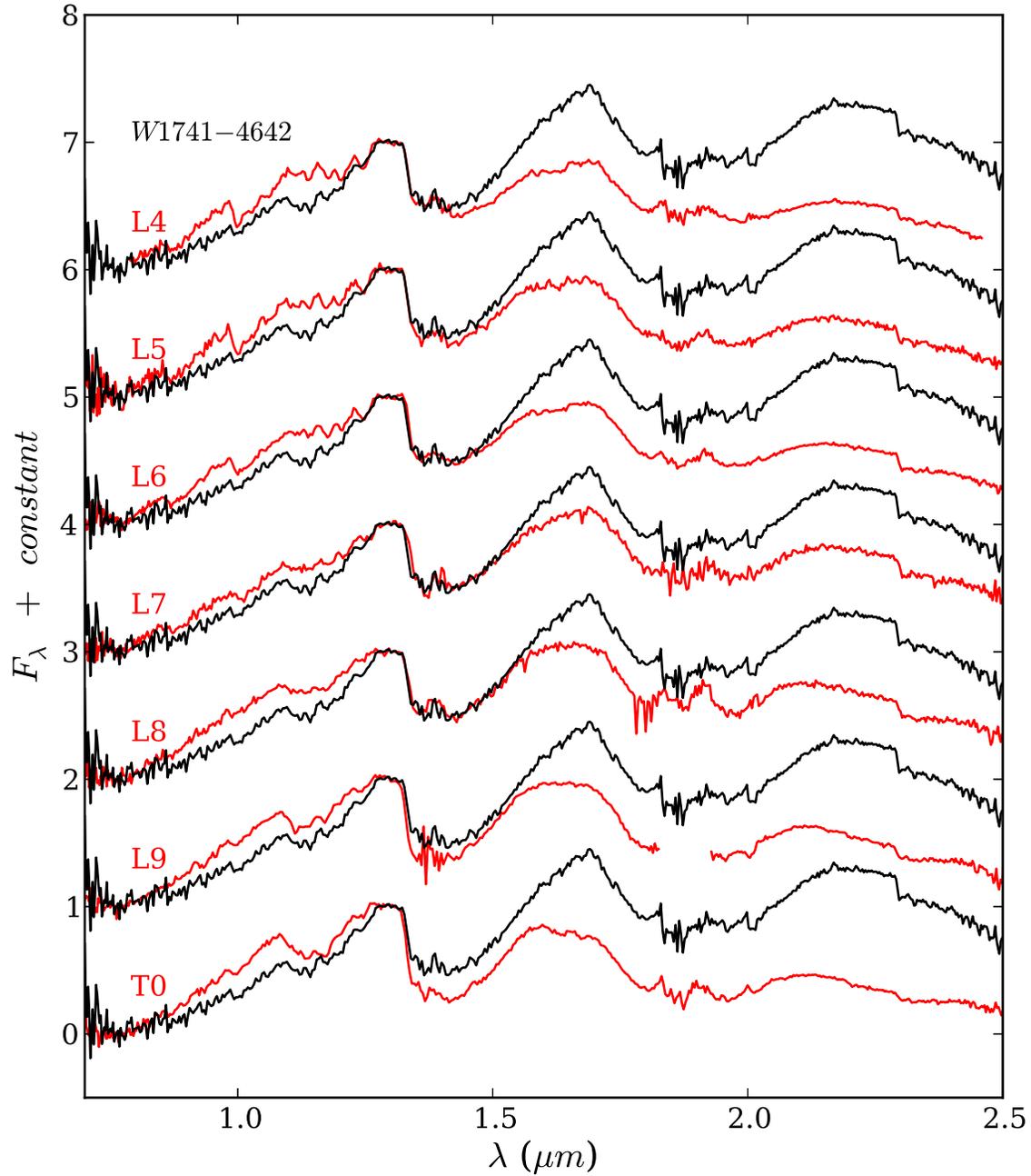}
\caption{The near-infrared spectrum of WISE 1741$-$4642 (black) compared with spectral standards (L4 through T0 - red) from the Spex Prism Spectral Library.  Each spectrum is normalized by the mean flux from 1.27 to 1.32 $\mu$m.  The spectra are separated along the vertical axis by a constant of 1.0 for clarity.  The spectral standards used for this comparison are as follows: 2MASS J21580457-1550098 (L4; \citealt{kirk10}), SDSS J083506.16$+$195304.4 (L5; \citealt{chiu06}), 2MASSI J1010148$-$040649 (L6; \citealt{reid06}), 2MASSI J0103320$+$193536 (L7; \citealt{cruz04}), 2MASSW J1632291$+$190441 (L8; \citealt{burg07}), DENIS-P J0255$-$4700 (L9; \citealt{burg06}), and SDSS J120747.17$+$024424.8 (T0; \citealt{loop07}).}
\end{figure}

\begin{figure}
\plotone{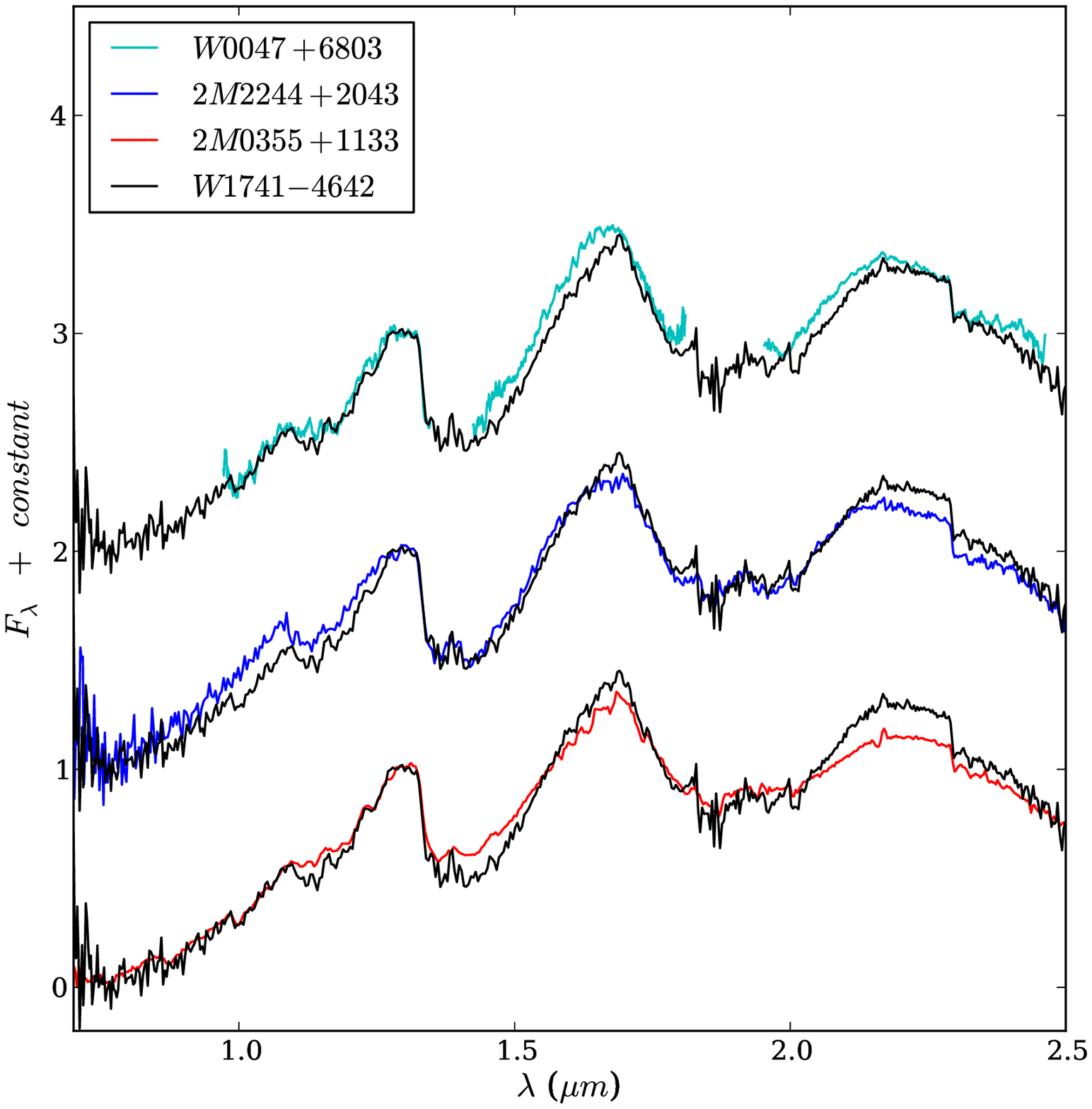}
\caption{The near-infrared spectrum of WISE 1741$-$4642 (black) compared with the unusually red L dwarfs W0047$+$6803 (SpT$_{IR}$ = L7.5 (pec) - \citealt{gizis12}), 2M2244$+$2043 (SpT$_{IR}$ =  L7.5 (pec) - \citealt{kirk08}), and 2M0355$+$1133 (SpT$_{opt}$ = L5$\gamma$ - \citealt{fah13}). Each spectrum is normalized by the mean flux from 1.27 to 1.32 $\mu$m.  While the spectra match better than the spectral standards in Figure 2, WISE 1741$-$4642 is redder than both 2M0355$+$1133 and 2M2244$+$2043 and more sharply peaked at H than 2M2244$+$2043.}
\end{figure}

\begin{figure}
\epsscale{0.8}
\plotone{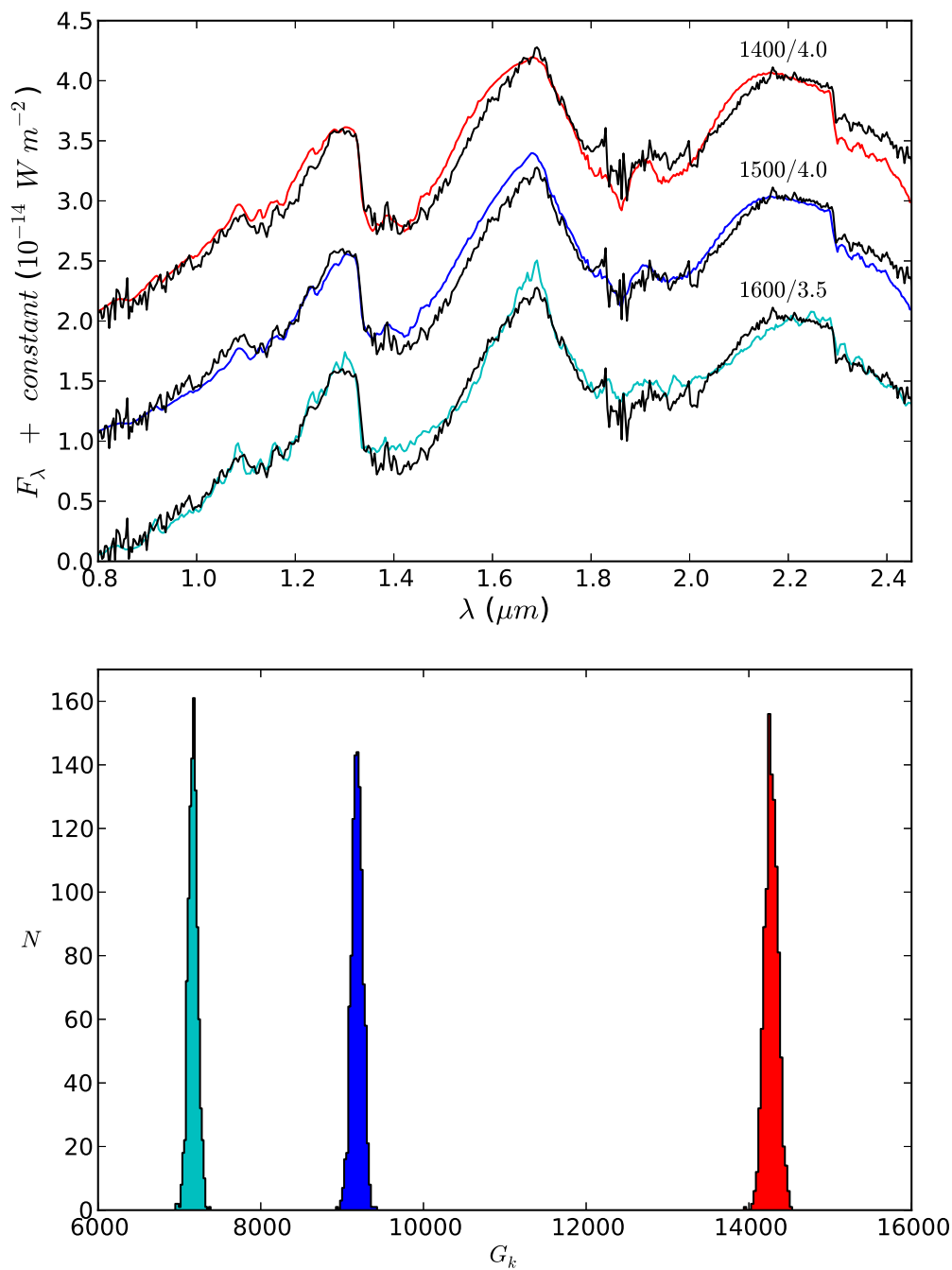}
\caption{{\it Top:} The near-infrared spectrum of WISE 1741$-$4642 (black) compared with the three best fitting BT-Settl models.  Each model spectrum is labeled with its corresponding $T_\mathrm{eff}$ and log $g$ values. {\it Bottom:} The distribution of $G_k$ values for the three best fitting models.  Red: $T_\mathrm{eff}$ = 1400 K and log $g$ = 4.0, blue: $T_\mathrm{eff}$ = 1500 K and log $g$ = 4.0, and cyan: $T_\mathrm{eff}$ = 1600 K and log $g$ = 3.5.}
\end{figure}

\begin{figure}
\epsscale{1.0}
\plotone{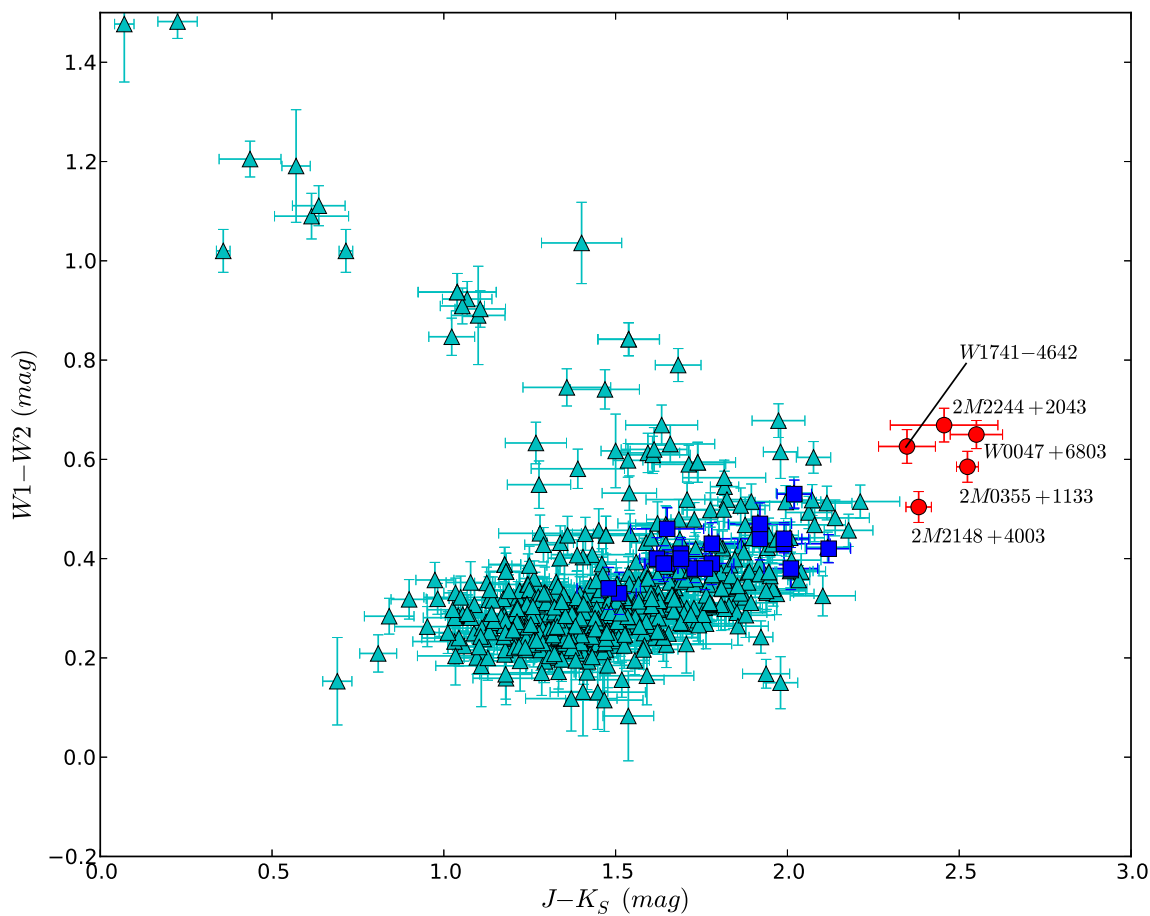}
\caption{2MASS J$-$K$_S$ vs.\ WISE W1$-$W2 colors for dwarfs with photometric uncertainties $\leq$ 0.1 mag from dwarfarchives.org (cyan triangles).  WISE 1741$-$4642 and red L dwarfs 2M2148+4003, 2M2244+2043, 2M0355+1133, and W0047+6803 are shown as a red circles.  Blue squares represent L$\gamma$ dwarfs from Table 3 of \cite{fah13}, the majority of which have optical spectral types earlier than L5. }
\end{figure}

\begin{figure}
\epsscale{1.0}
\plotone{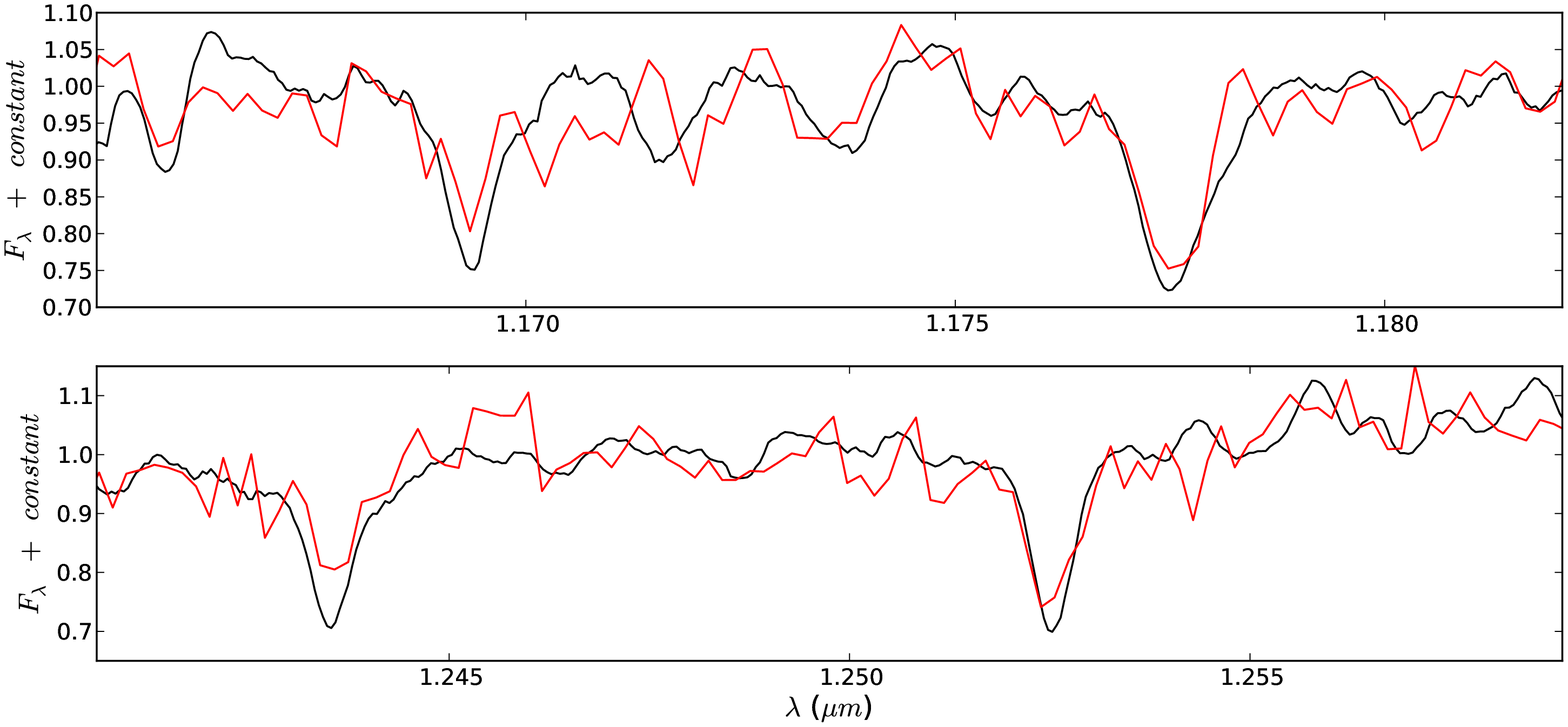}
\caption{A K I line comparison of the WISE 1741$-$4642 FIRE spectrum (black) with the red L7.5 dwarf W0047$+$6803 (red) \citep{gizis12}.  The FIRE spectrum was smoothed with a gaussian kernel to the resolution of the comparison spectrum. The flux of both spectra were normalized to the continuum.}
\end{figure}

\begin{figure}
\plotone{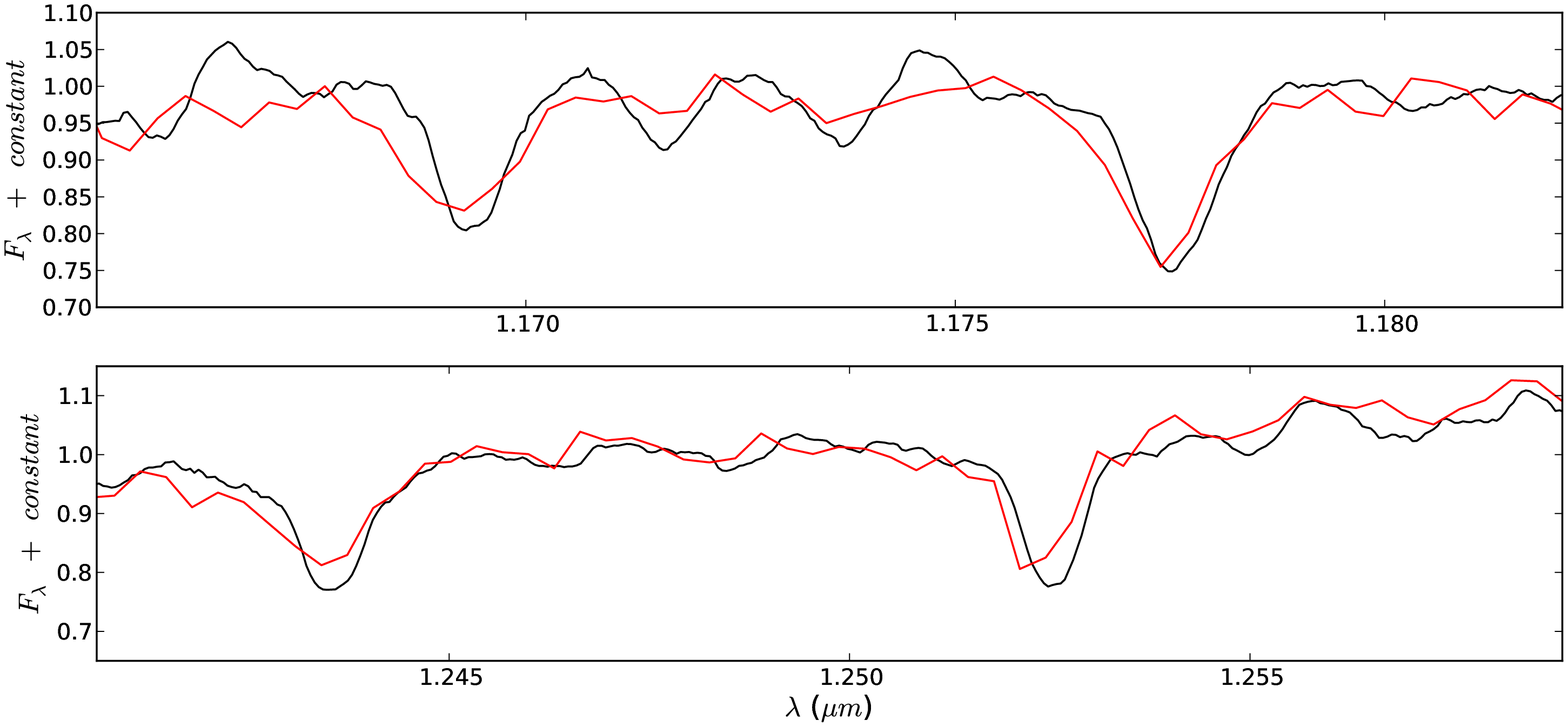}
\caption{A K I line comparison of the WISE 1741$-$4642 FIRE spectrum (black) with the L5$\gamma$ dwarf 2M0355$+$1133 (red) \citep{fah13}.  The FIRE spectrum was smoothed with a gaussian kernel to the resolution of the comparison spectrum. The flux of both spectra were normalized to the continuum.}
\end{figure}

\begin{figure}
\plotone{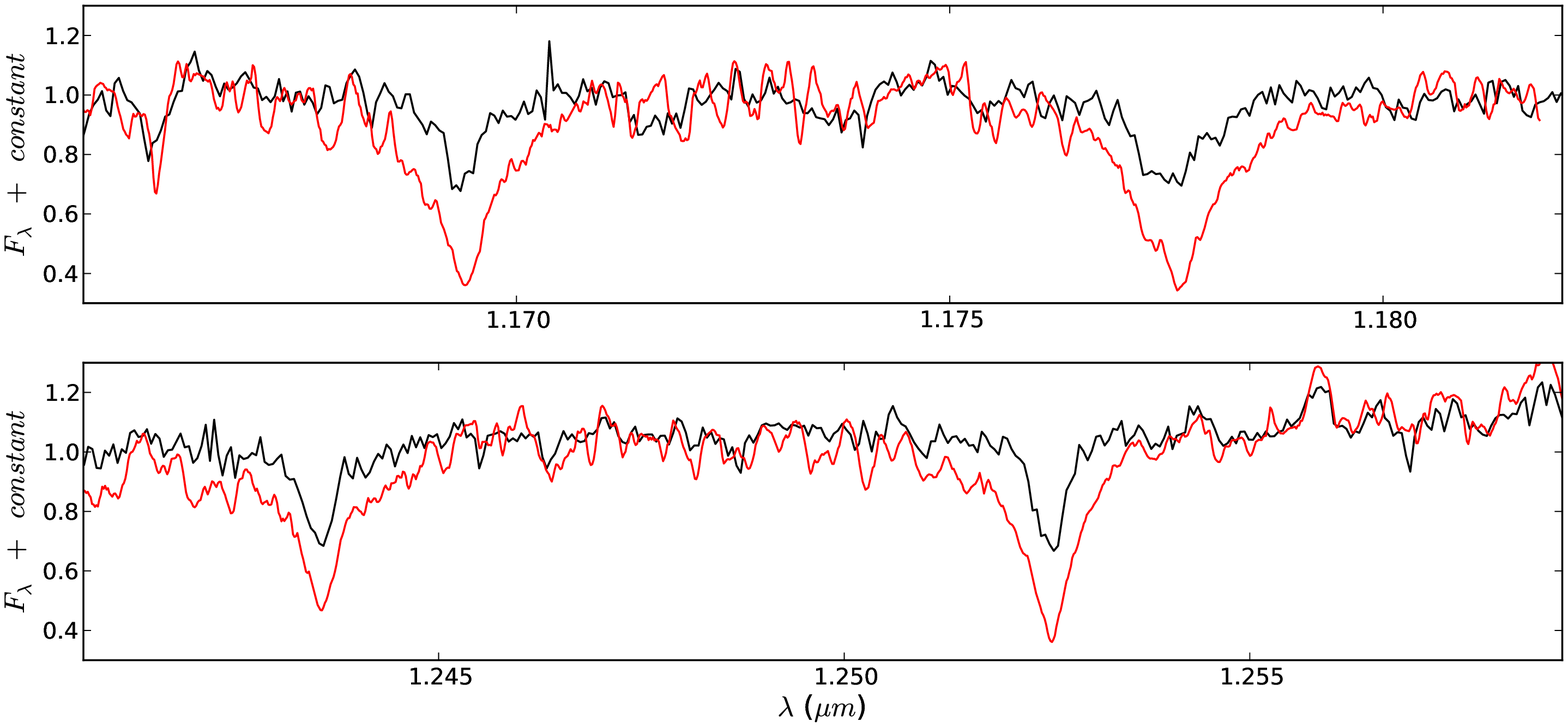}
\caption{A K I line comparison of the WISE 1741$-$4642 FIRE spectrum (black) with the L5 dwarf 2M1507$-$16 spectrum from the BDSS (red) \citep{mcl07}.  The NIRSPEC spectrum was smoothed with a gaussian kernel to the resolution of the FIRE spectrum. The flux of both spectra were normalized to the continuum.}
\end{figure}

\begin{figure}
\plotone{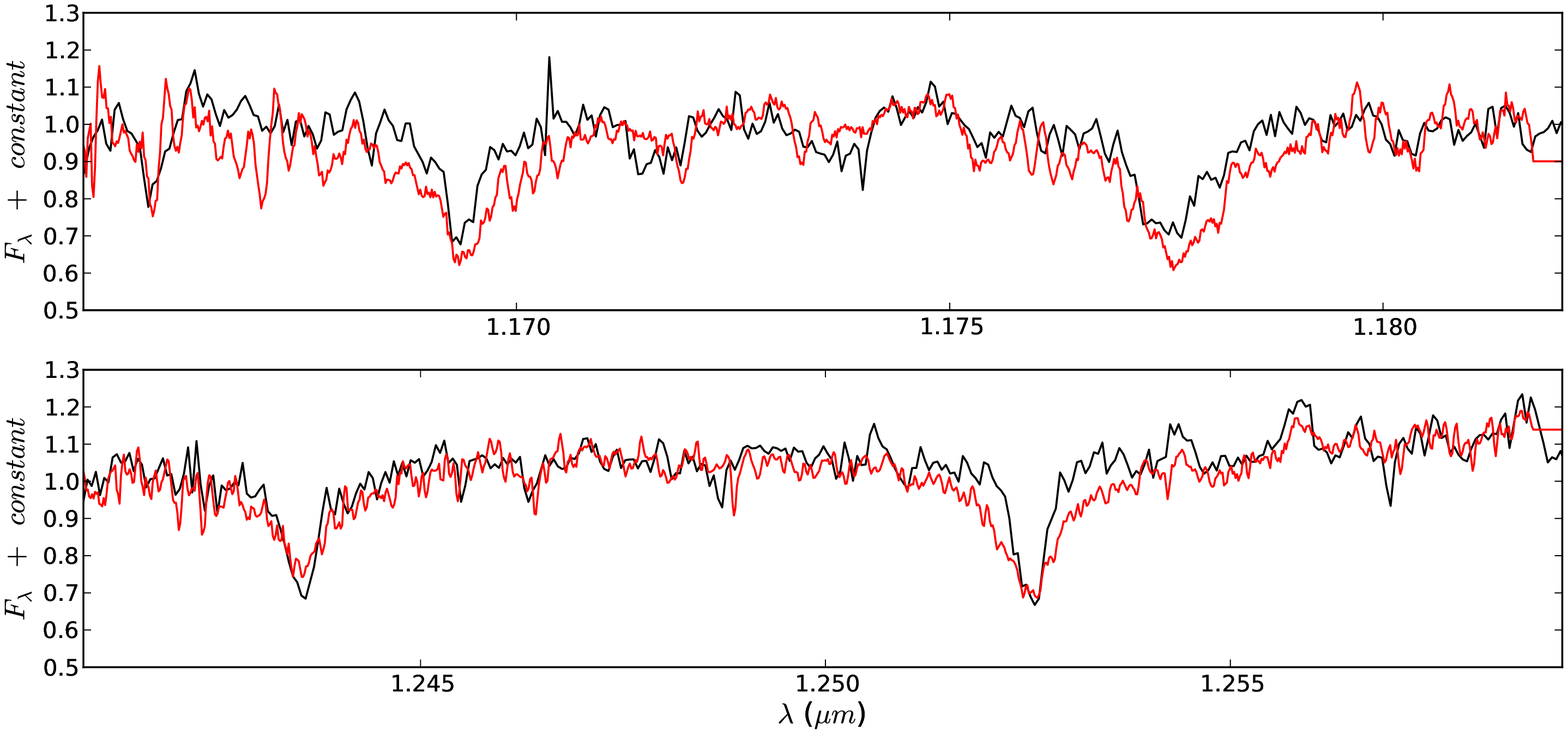}
\caption{A K I line comparison of the WISE 1741$-$4642 FIRE spectrum (black) with the L7 dwarf DENIS0205$-$11 spectrum from the BDSS (red) \citep{mcl07}.  The NIRSPEC spectrum was smoothed with a gaussian kernel to the resolution of the FIRE spectrum. The flux of both spectra were normalized to the continuum.}
\end{figure}

\begin{figure}
\plotone{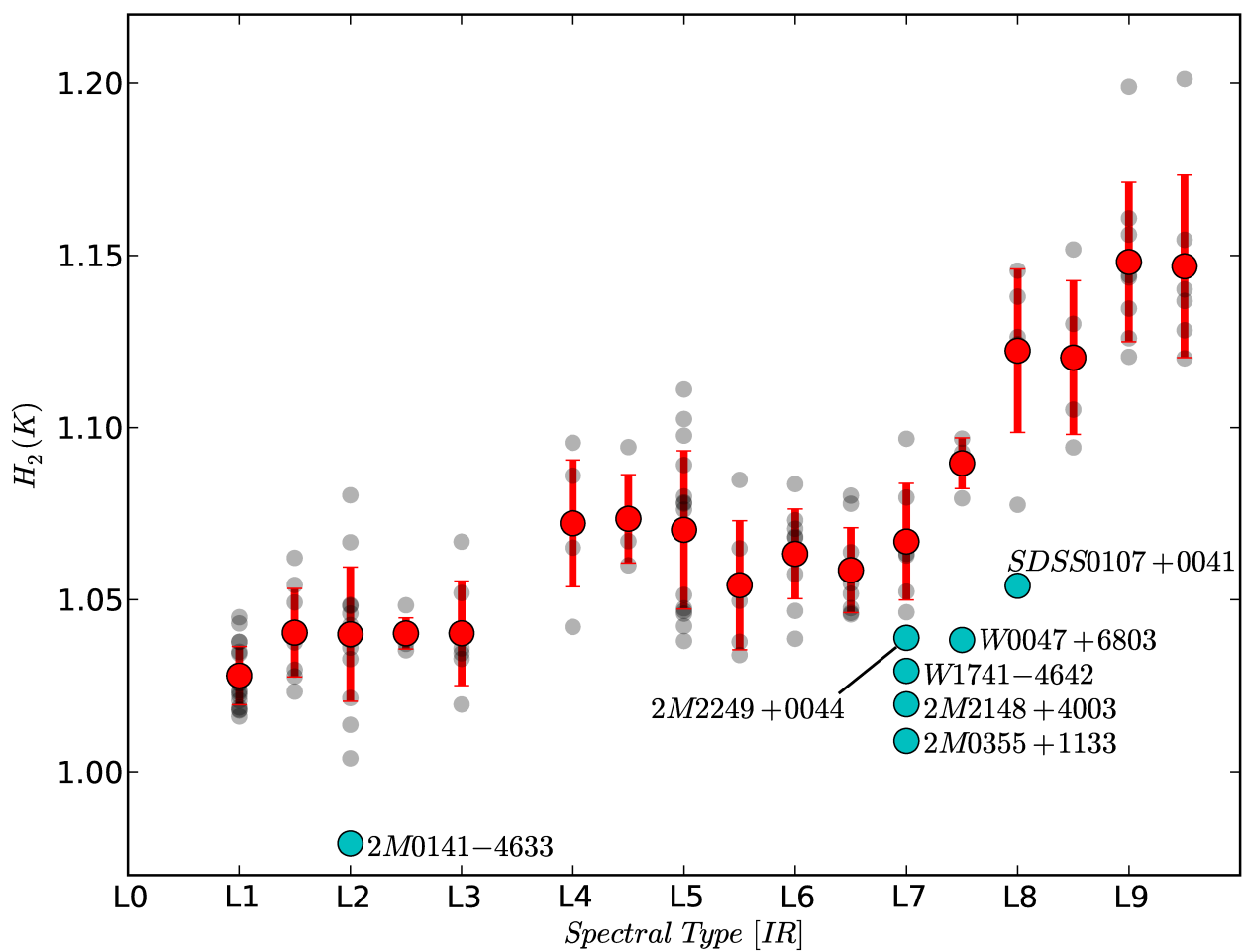}
\caption{$H_2$$(K)$ index as a function of spectral type for L dwarfs from the SpeX Prism Spectral Library based on our new spectral typing system.  Red points indicate the mean $H_2$$(K)$ index and standard deviation for L dwarfs binned by 0.5 subtype.  The young and/or red L dwarfs 2M2148$+$4003, 2M0355$+$1133, 2M0141$-$4633, 2M2249$+$0044, W0047$-$6803, and SDSS 0107$+$0041, as well as WISE 1741$-$4642 are individually labeled.  }
\end{figure}

\begin{figure}
\epsscale{0.6}
\plotone{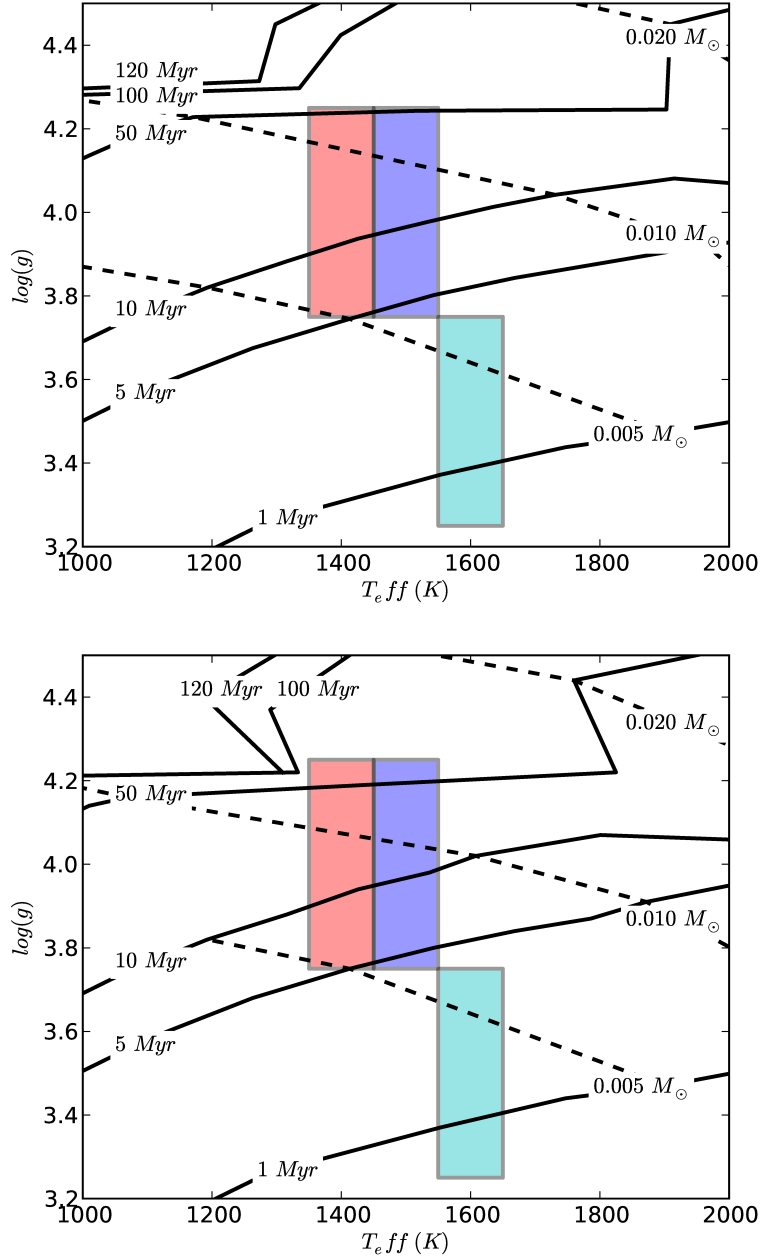}
\caption{{\it Top:} Theoretical ``COND'' log $g$ vs.\ $T_\mathrm{eff}$ evolutionary tracks from \cite{bar03}.  1, 5, 10, 50, 100, and 120 Myr isochrones are shown by solid lines from right to left.  Curves of constant mass from 0.005, 0.010, and 0.020 M$_{\sun}$ (5, 10, and 21 M$_{Jup}$) are shown by dashed lines from left to right.  The red, blue, and cyan shaded regions represent the three best model spectrum fits; $T_\mathrm{eff}$ = 1400 K and log $g$ = 4.0 (red), $T_\mathrm{eff}$ = 1500 K and log $g$ = 4.0 (blue), $T_\mathrm{eff}$ = 1600 K and log $g$ = 3.5 (cyan).  The colors match those in Figure 6.  {\it Bottom:} Same figure as on the {\it Left} for the theoretical ``DUSTY'' log $g$ vs.\ $T_\mathrm{eff}$ evolutionary tracks from \cite{cha00}. 
}
\end{figure}

\begin{figure}
\epsscale{0.6}
\plotone{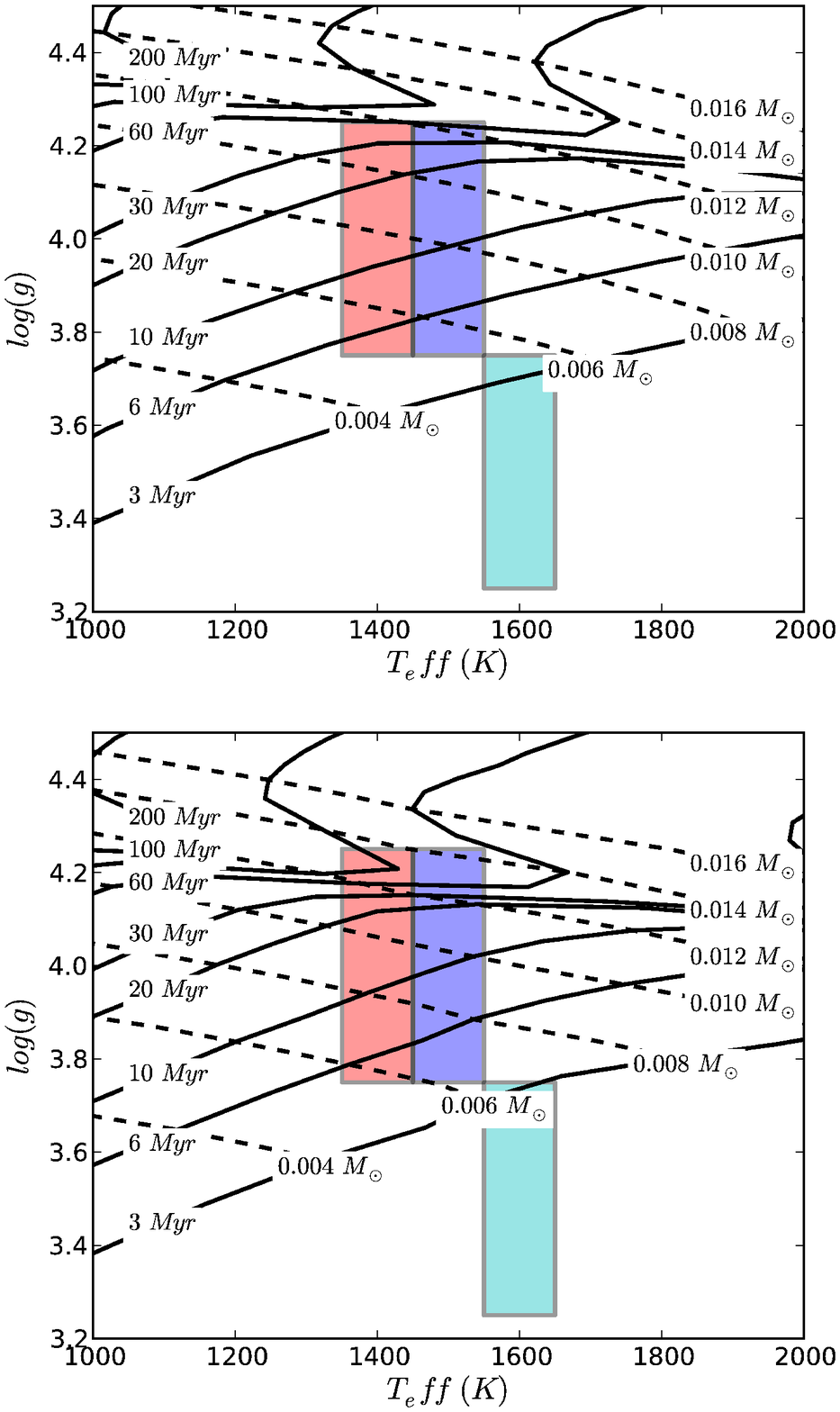}
\caption{{\it Top:} Theoretical (cloudless) log $g$ vs.\ $T_\mathrm{eff}$ evolutionary tracks from \cite{sau08}.  3, 6, 10, 20, 30, 60, 100, and 200 Myr isochrones are shown by solid lines from right to left.  Curves of constant mass from 0.004$-$0.016 M$_{\sun}$ (4$-$17 M$_{Jup}$) in steps of 0.002 M$_{\sun}$ are shown by dashed lines from left to right.  The red, blue, and cyan shaded regions represent the three best model spectrum fits; $T_\mathrm{eff}$ = 1400 K and log $g$ = 4.0 (red), $T_\mathrm{eff}$ = 1500 K and log $g$ = 4.0 (blue), $T_\mathrm{eff}$ = 1600 K and log $g$ = 3.5 (cyan). The colors match those in Figure 6.  {\it Right:} Same figure as on the {\it Bottom} for the theoretical (cloudy - f$_{sed}$ = 2) log $g$ vs.\ $T_\mathrm{eff}$ evolutionary tracks from \cite{sau08}.  
}
\end{figure}

\begin{figure}
\epsscale{0.6}
\plotone{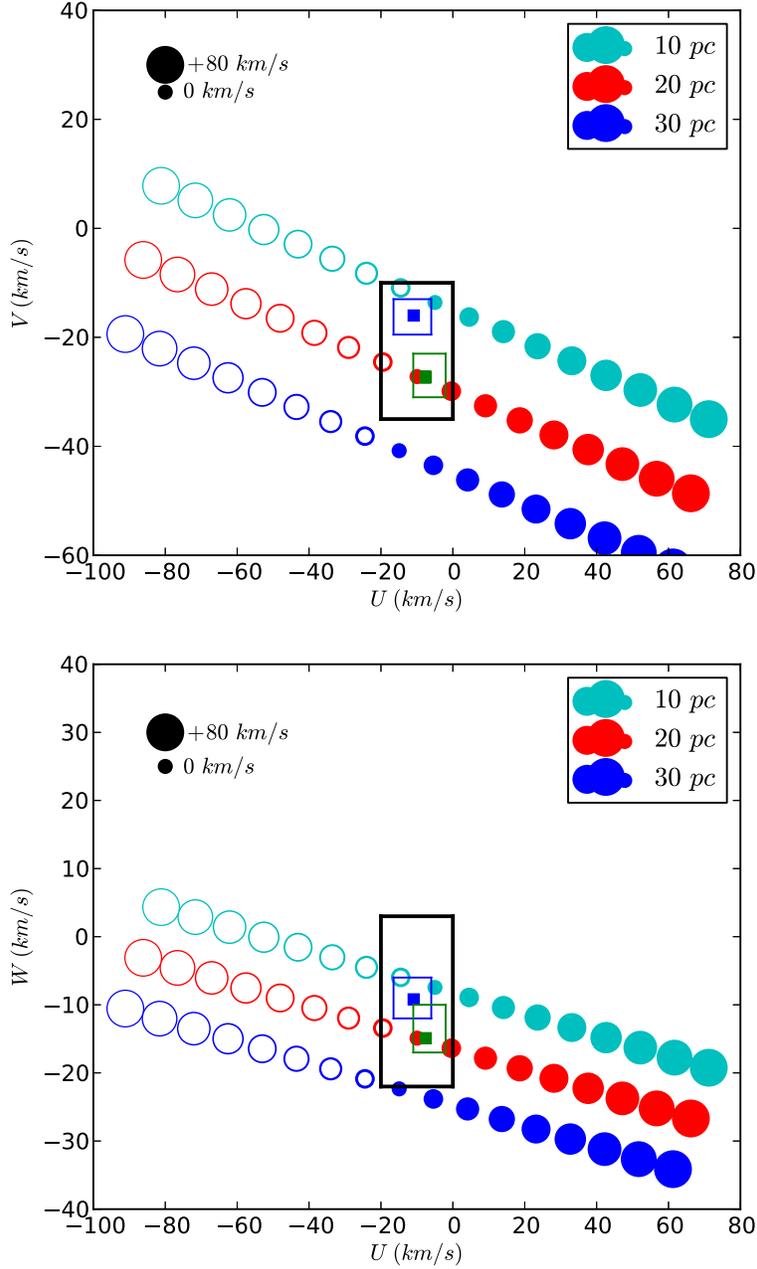}
\caption{UVW space velocities for WISE 1741$-$4642 for a range of radial velocities (-80 to +80 km s$^{-1}$ in 10 km s$^{-1}$ steps) for distances of 10, 20, and 30 pc.  Negative radial velocity values are represented by open circles.  The black square indicates the ``good box'' from \cite{zuck04}.  The blue and green empty boxes indicate the extent of highly probably members from \cite{torres08} for $\beta$ Pictoris and the AB Dor association, respectively.  Small blue and green filled squares represent the core UVW values of $\beta$ Pictoris and AB Doradus as defined in \cite{fah13}.   
}
\end{figure}

\begin{figure}
\epsscale{0.7}
\plotone{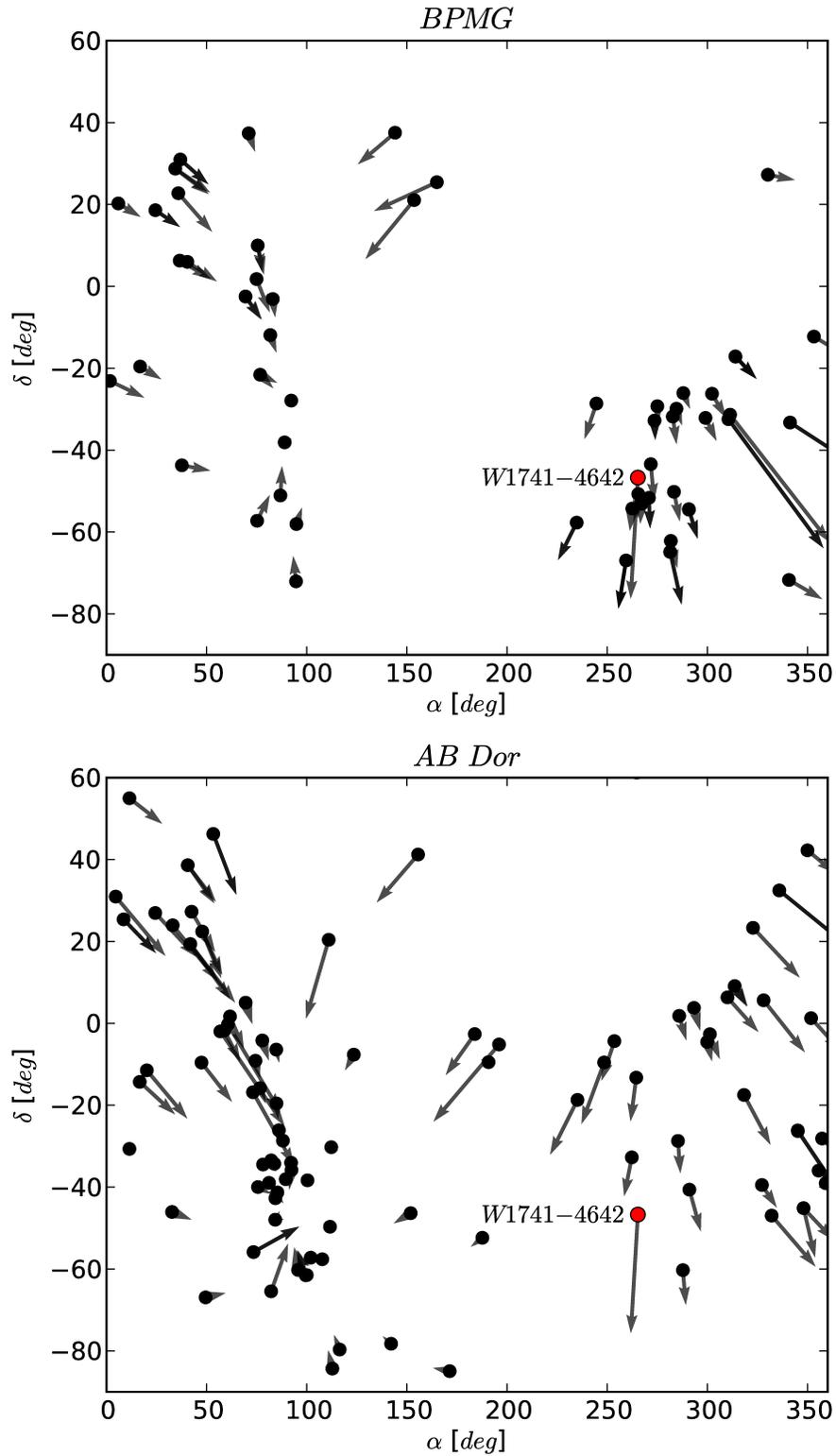}
\caption{{\it Left:} The positions and proper motion of WISE 1741$-$4642 along with members of the Beta Pictoris moving group.  {\it Right:} The positions and proper motion of WISE 1741$-$4642 along with members of the AB Doradus moving group.   
}
\end{figure}

\begin{figure}
\epsscale{0.7}
\plotone{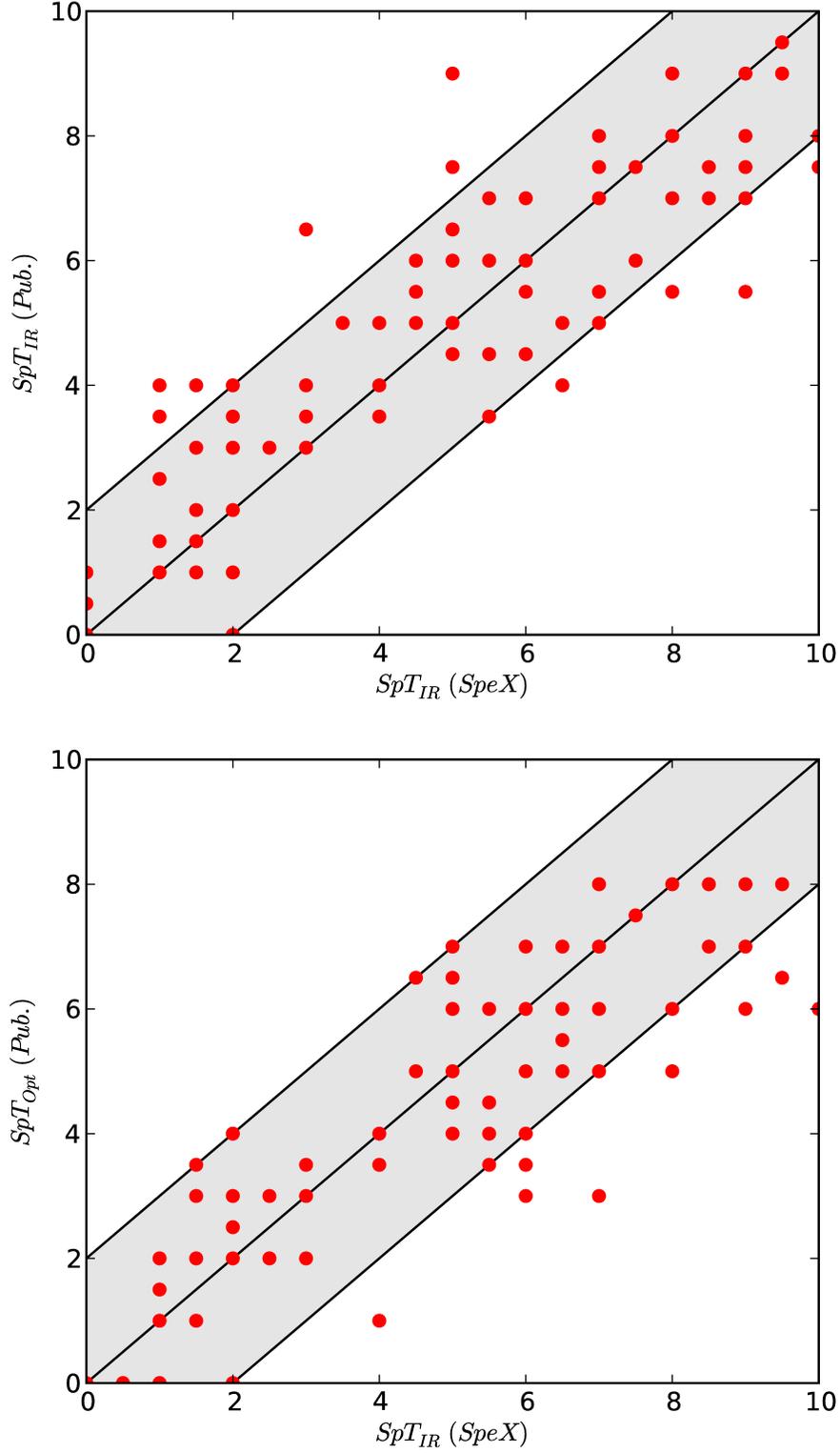}
\caption{{\it Left:} A comparison of the spectral types derived in Appendix A and presented in Table 2 with previously published near-infrared spectral types.  Any point in the shaded region agrees to within $\pm$2 subtype.  {\it Right:}  A comparison of the spectral derived in Appendix A and presented in Table 2 with previously published optical spectral types.  We note that there are no optical spectral types defined between L8 and T0.}
\end{figure}

\end{document}